\documentclass[a4paper,fleqn,usenatbib]{mnras}
\usepackage{lineno}
\usepackage[T1]{fontenc}
\usepackage{ae,aecompl}

\usepackage{graphicx}	% Including figure files
\usepackage{amsmath}	% Advanced maths commands
\usepackage{amssymb}	% Extra maths symbols
\usepackage{subfig}
\usepackage{hyperref}
\usepackage{soul}
\title[Spectral study of Mkn\,421 using Swift-XRT observations]{Correlations between X-ray spectral parameters of Mkn\,421 using long-term Swift-XRT data}

\author[R. Khatoon et al.]{
Rukaiya Khatoon$^{1,2}$\thanks{E-mail: rukaiyakhatoon12@gmail.com},
Zahir Shah$^{1,3}$,
Jyotishree Hota$^{4}$,
Ranjeev Misra$^{1}$,
\newauthor Rupjyoti Gogoi$^{2}$, 
and Ananta C. Pradhan$^{4}$
\\
$^{1}$Inter-University Center for Astronomy and Astrophysics, Post Bag 4, Ganeshkhind, Pune-411007, India. \\
$^{2}$Tezpur University,Napaam-784028, Assam, India.\\
$^{3}$Department of Physics, Central University of Kashmir, Ganderbal 191201, India.\\
$^{4}$Department of Physics and Astronomy, National Institute of Technology, Rourkela, Odisha 769008, India.
}

\date{Accepted XXX. Received YYY; in original form ZZZ}

\pubyear{2020}

\begin{document}
\label{firstpage}
\pagerange{\pageref{firstpage}--\pageref{lastpage}}
\maketitle

\begin{abstract}

We have performed a detailed analysis of the X-ray spectra of the blazar Mkn\,421 using Swift-XRT observations taken  between 2005 and 2020, to quantify the correlations between spectral parameters for different models. In an earlier work, it has been shown that such spectral parameter correlations obtained from a single short flare of duration $\sim$ 5-days of Mkn\,421, can be used to distinguish spectrally degenerate models and  provide estimates of physical quantities. In this work, we show that the results from the long-term spectral parameter correlations are consistent with those obtained from the single flare. In particular,  that the observed spectral curvature is due to maximum cutoff energy in the particle distribution is ruled out. Instead, models where the curvature is due to the energy dependence of escape or acceleration time-scale of the particles are favored. The estimated values of the physical parameters for these models are similar to the ones obtained from the single flare analysis and are somewhat incompatible with the physical assumption of the models, suggesting that more complex physical models are required. The consistency of the results obtained from the long and short-term evolution of the source, underlines the reliability of the technique to use spectral parameter correlations to distinguish physical models. 

%We perform the correlation study between the model fit parameters and the results show .  
\end{abstract}


%%%%%%%%%%%%%%%%% BODY OF PAPER %%%%%%%%%%%%%%%%%%

\section{Introduction}
%####################### 
Blazars are the jetted Active Galactic Nuclei (AGNs) whose relativistic jet is aligned close to our line of sight \citep{1993ARA&A..31..473A, 1995PASP..107..803U}. They show rapid flux variations across the entire accessible electromagnetic spectrum at all time scales from minutes to years \citep{doi:10.1146/annurev.astro.35.1.445}. They are also characterized by strong radio and optical polarization, apparent superluminal motion in high resolution radio maps. Blazars are classified into two subgroups, namely: BL-Lac objects, which have very faint optical emission lines \citep{10.1093/mnras/281.2.425}, and the Flat Spectrum Radio Quasars (FSRQs), which show strong emission features in the optical spectra \citep{1995PASP..107..803U, 2003ApJ...585L..23F}.

The Spectral Energy Distribution (SED) of blazars has a double hump structure \citep{1997A&A...327...61G}, which is dominated by the emission from the beamed relativistic jets. The low energy component peaking in the IR - X-ray regime, is well known to be caused by the synchrotron emission from the ultra-relativistic electrons spiraling around the magnetic field lines in the jet \citep{1995PASP..107..803U}. While, the high energy component peaks at MeV-TeV gamma-ray energies, the processes responsible for emission are still unresolved. Largely, models including leptonic and hadronic are assigned to explain the origin of the high energy emission \citep{1995ApJ...441...79B, 2000A&A...353..847A}. In case of the leptonic model, the inverse Compton (IC) scattering of the low energy photons can reproduce the high energy hump \citep{1992A&A...256L..27D, 1993ApJ...416..458D, 2017MNRAS.470.3283S}. The low energy photons can be either synchrotron photons or the photons from sources that are external to jet. In the former case, the IC process is called Synchrotron Self Compton (SSC) \citep{1985ApJ...298..114M}; whereas the latter one is termed as External Compton (EC) \citep{1992A&A...256L..27D}. The hadronic model explains the high energy component by the emission through relativistic protons, and the process can be proton-synchrotron interactions \citep{2011ApJ...736..131A} or the  proton-photon interactions \citep{1993A&A...269...67M}. Depending on the position of the peak frequency in the synchrotron component, BL Lacs are further classified into three subgroups viz. low synchrotron-peaked BL\,Lac (LBL; $\nu_{p} {\leq} 10^{14}$ Hz) \citep{2016ApJS..226...20F}, intermediate synchrotron-peaked BL\,Lac (IBL; $10^{14} < \nu_{p} {\leq} 10^{15.3}$ Hz), and high synchrotron-peaked BL\,Lac (HBL; $\nu_{p}>10^{15.3}$ Hz; \citep{Abdo_2010}). 

Mkn\,421 is the nearest (z = 0.031) HBL type object, with the peak frequency of the low-energy component beyond $10^{15.3}$ Hz. It is the brightest BL Lac object in UV and X-ray bands, and was first detected in $\gamma$-rays using the Energetic Gamma Ray Experiment Telescope (EGRET) \citep{1992ApJ...401L..61L}. It was also the first blazar observed at TeV energies using the Whipple telescope \citep{1992Natur.358..477P, 1996A&A...311L..13P}, and thus classified it as a TeV blazar. The object shows rapid flux variations in multi-wavebands which are often characterized by the spectral changes in the emitting electrons \citep{2000ApJ...541..153F, 2004A&A...413..489M, 2006ApJ...647..194X, B_ttcher_2010, 2014JApA...35..241G}. Several X-ray observations suggest that the X-ray spectrum exhibits a well-marked curvature during both flaring and quiescent states. The observed curved spectrum is found to be well fitted by the log-parabola function \citep{1986ApJ...308...78L, 2004ApJ...601..759T, 2004A&A...413..489M, 2007A&A...467..501T, 2009A&A...504..821P, 10.1093/mnras/stw095, Gaur_2017, 2018MNRAS.480.2046G, 2018A&A...619A..93B, 2018ApJ...859...49P}.  

The log-parabola shape in the photon spectrum can be explained with the log-parabola particle distribution and a possible signature of producing such curved electron distribution was found in terms of acceleration processes. \citealp{2004A&A...413..489M, 2006A&A...448..861M} interpreted such feature in the particle distribution by considering the acceleration probability to be energy dependent such that it decreases with the electron energy. They performed a detailed X-ray spectral analysis using BeppoSAX, Swift, and XMM-Newton observations during 1996 -- 2007. Their study showed that a linear relation between the curvature parameter ($\beta$) and the spectral index ($\alpha$), supports the hypothesis that the log-parabolic curve can be produced due to the particle acceleration mechanism. Such positive correlation between $\alpha$ and $\beta$, can result from the shock acceleration \citep{2004A&A...413..489M}. Subsequently, \citealp{2017ApJ...848..103K, 2018ApJ...854...66K, 2020ApJS..247...27K} found such positive correlation between $\alpha$ and $\beta$ during the XRT-observations in 2005 -- 2008 and 2015 -- 2018.  Furthermore, an anticorrelation between the synchrotron peak energy ($E_p$) and the curvature ($\beta$) indicates that the log-parabola particle distribution can be associated with the energy dependent acceleration mechanism, while curvature decreases as the acceleration becomes more efficient \citep{2004A&A...413..489M}. Additionally, several studies show that $E_p$ is positively correlated with the flux in the energy range 0.3 -- 10 keV ($F_{0.3-10keV}$) \citep{2007A&A...467..501T, 2008A&A...478..395M, 2009A&A...504..821P}, indicating a shift of the peak to higher energies with increasing flux. An alternate interpretation is provided by the stochastic acceleration of the electrons with the magnetic turbulence \citep{2007A&A...467..501T, 2009A&A...501..879T, 2020ApJS..247...27K}. Moreover, \cite{1962SvA.....6..317K, 1969Ap&SS...5..131M} and \cite{2011ApJ...739...66T} demonstrated that log-parabolic spectral distributions are naturally produced in the framework of stochastic acceleration, i.e. when a diffusion in momentum space is included.

Alternative to the log-parabola model, curvatures in the particle spectrum may arise when the escape time-scale from the emission region is energy dependent; while, the escape and acceleration time-scales in acceleration region (AR) are energy independent \citep{Sinha_2017}. In different circumstances, broken log-parabola model considering energy dependent escape timescale from AR, is also capable of reproducing the curved spectrum \citep{10.1093/mnrasl/sly086}.
%Alternative to the empirical model, analytical models have also been used to study the curvature and its physical significance. 
Furthermore, \citealp{2018MNRAS.480.2046G} found that, energy dependent electron diffusion (EDD) model is also capable to describe the X-ray broad-band SED for Mkn\,421, assuming escape rate of electrons from acceleration region is energy dependent. They showed that EDD model fit for the simultaneous NuSTAR and Swift-XRT observations is marginally better than the log-parabola one, and also this model gives a better understanding about the emission processes in the jet \citep{2018MNRAS.480.2046G}. Furthermore, \citealp{2020MNRAS.492..796G} showed that the power-law with $\gamma_{max}$ model can explain the curved spectrum in the X-ray band in case of EHBL source RGB J0710+591, where the synchrotron spectrum corresponding to the decline of the particle number density near the maximum available electron energy, can explain the sharp spectral curvature. 
%A simplistic approach is considered in \citealp{2021MNRAS.tmp.2632H}, where curved spectrum can be interpreted as an outcome of energy dependent acceleration and diffusion of the particles.
Nevertheless, \citealp{2021MNRAS.tmp.2632H} had shown with the approach of spectral parameter correlations that the obtained correlations are not consistent with the predictions from the power-law with $\gamma_{max}$ model. Instead, they showed that curved spectrum can be interpreted as an outcome of energy dependent acceleration and diffusion of the particles. 

In particular, \emph{Swift}-XRT carried out an extensive study of Mkn\,421 at X-ray (0.3 -10 keV) energy band on different epochs. The extreme X-ray brightness of the source, allows us to carry out a detailed spectral study even with the exposure of a few hundred seconds. Several authors presented an extensive study
of Mkn\,421 by using the XRT data during the periods between 2005 March - 2018 April \citep{Kapanadze_2016, 10.1093/mnras/stw095, 2017ApJ...848..103K, 2018ApJ...854...66K, 2018ApJ...858...68K, 2020ApJS..247...27K}. Their study revealed strong X-ray flares brighten by a factor of 3-20 on the time-scale of a few days to weeks. They further showed that the spectra were fitted well with the log-parabola model for the flaring states, showing relatively low curvature which can be described by the stochastic acceleration of particles. 
%if the 0.3-10 keV X-ray spectra observed with Swift-XRT, consisting both flaring and quiescent states are described well with the empirical models or the analitycally motivated models. 
In this study, we present for the first time, the spectral study of Mkn\,421 using an entire collection of Swift-XRT data available during the period April 2005 to April 2020. We fit each spectrum with different particle energy distribution models viz. log-parabola model, power-law particle distribution with maximum electron energy, energy-dependent diffusion (EDD), and energy-dependent acceleration (EDA) models, and the corresponding results are discussed. We perform a correlation study between the best fit model parameters and compare them with the ones obtained from the  log-parabola spectral fit. \citealp{2021MNRAS.tmp.2632H}, used these models and showed that the correlations between the spectral parameters for a single short-term flare with a duration $\sim$ 400 ksec can be used to constraint different models and provide an assessment of the underlying physical quantities. The motivation of this work is to test whether the correlation results from the long-term observations are consistent with the one obtained from short-term flare \citep{2021MNRAS.tmp.2632H}. 

The paper is assembled as follows: in \S 2, we describe about the data analysis procedure. In \S 3, we present the spectral study using log-parabola model (\S 3.1), Power-law with maximum electron energy model (\S 3.2), and Energy dependent model (\S 3.3); EDD (\S 3.3.1) and EDA (\S 3.3.2). In \S 3.4, we compare the correlation study results with the ones obtained with short-term flare in \citep{2021MNRAS.tmp.2632H}. Finally, we discuss and conclude the results in \S 4.    

\section{Swift-XRT observation}

NASA's HEASARC interface \footnote{https://heasarc.gsfc.nasa.gov/} provides the \emph{Swift}-XRT data in the 0.3 -- 10 keV band. We have performed here the spectral study of Mkn\,421, using an entire collection of the \emph{Swift}-XRT observations, during the period from April 2005 to April 2020. There are 1117 pointings available during this period. The spectra were obtained using the on-line \emph{Swift}-XRT products generation tool \footnote{https://www.swift.ac.uk/user\_objects/} \citep{2009MNRAS.397.1177E}, which uses HEASOFT software version 6.22. The tool creates the X-ray light curves, spectra, and images of any point source, which falls in the \emph{Swift}-XRT field of view, and the products are corrected from the instrumental artifacts i.e., pile up and the bad column CCD corrections. The events with 0--2 grades performed in the Windowed Timing (WT) mode has been considered in the analysis. The XRT online tool provides 0.3 -- 10 keV X-ray spectra with one count in each bin. We used the grppha tool to obtain 20 counts in each bin, in order to make the spectrum valid for the ${\chi}^{2}$-statistics. Further, in order to obtain constrained spectral parameters we did not consider 32 number of observations for which the spectral counts of each observation was < 3000. However, we have verified that the qualitative results presented in this work do not change when these observations are included.\\

\section{Spectral analysis}
The X-ray spectrum of Mkn\,421 is known to be produced by the non-thermal relativistic electrons undergoing synchrotron emission. Hence we model the non-thermal X-ray emission from Mkn\,421 by assuming that the emission originates from a spherical region of radius, R. The emission region is assumed to be filled with tangled magnetic field, B and relativistic isotropic electron distribution, n($\gamma)$ which undergoes synchrotron loss.
The synchrotron emissivity due to a relativistic electron distribution n($\gamma)$ can be estimated by using the equation 
\begin{align}\label{eq:syn_emiss}
	J_{\rm syn}({\epsilon}')= \frac{1}{4\pi}\int P_{\rm syn}(\gamma,{\epsilon}')\,n(\gamma)\,d\gamma
\end{align}
here, $P_{\rm syn}(\gamma,{\epsilon}')$\footnote{In this paper, the $'$ indicates that the physical quantity is estimated in the emission region frame.} is the pitch angle averaged synchrotron power emitted by single particle and can be obtained using the equation \citep{1986rpa..book.....R}

\begin{align}\label{eq:syn_power}
	P_{\rm syn}(\gamma, \epsilon')= \frac{\sqrt{3} \pi e^3B}{4m_e c^2} f\left(\frac{\epsilon'}{\epsilon_c}\right)
\end{align}
where $\epsilon_c=\frac{3he\gamma^2B}{16m_ec}$ and $f\left(\frac{\epsilon'}{\epsilon_c}\right)$ is the synchrotron power function defined as \citep{1986rpa..book.....R}
\begin{align}\label{eq:syn_func}
	f(x)=x\int_{x}^{\infty} K_{5/3}(\psi)\,d\psi
\end{align}
with $K_{5/3}$ being the modified Bessel function of order 5/3.
Using the single particle synchrotron power (Equation \ref{eq:syn_power}) in Equation \ref{eq:syn_emiss}, and substituting $\xi=\gamma\sqrt{\mathbb{C}}$, where $\mathbb{C}=\frac{\delta}{1+z}\frac{3heB}{16m_ec}$ with z being the redshift of source and $\delta$ as jet Doppler factor, the synchrotron emissivity in the observed frame can be obtained as 
\begin{align}\label{eq:syn_emiss_1}
	J_{\rm syn}\left(\frac{1+z}{\delta}\epsilon\right)= \mathbb{A} \int_{\xi_{min}}^{\xi_{max}} f(\epsilon/\xi^2)n(\xi)d\xi
\end{align}
where $\mathbb{A}=\frac{\sqrt{3}\pi e^3B}{16m_ec^2\sqrt{\mathbb{C}}}$.  Finally, the
 synchrotron flux received by the observer at energy $\epsilon$ will be given by \citep{RevModPhys.56.255},\\
 \begin{align}\label{eq:obs_flux}
	F_{\rm syn}(\epsilon) &= \frac{\delta^3(1+z)}{d_L^2} V 
	J_{\rm syn}\left(\frac{1+z}{\delta}\epsilon\right) \nonumber\\
	&= \frac{\delta^3(1+z)}{d_L^2} V \,\mathbb{A} \int_{\xi_{min}}^{\xi_{max}} f(\epsilon/\xi^2)n(\xi)d\xi
\end{align}
where, $d_L$ is the luminosity distance, and V is the volume of the emission region.
This SYNchrotron CONVolution equation including single-particle emissivity and particle number density  is solved numerically and is used
%and the particle synchrotron emissivity is used to reproduce the particle synchrotron spectrum. The convolved equation is used
as a local convolution model \emph{($synconv \otimes n(\xi)$)} in XSPEC (Version 12.11.0) software package \citep{1996ASPC..101...17A}.  In the model, the  XSPEC ``energy" variable is represented as
$\xi = \sqrt{\mathbb{C}}\gamma$ such that the corresponding observed photon energy is $\epsilon=\xi^2$ \citep{2021MNRAS.tmp.2632H}. The \emph{($synconv \otimes n(\xi)$)} model outputs the synchrotron spectrum for a system with particle density, $n(\xi)$ as an input to the model. Also, it is established by various studies that the X-ray spectrum  exhibits mild curvature \citep{Fossati_2000, 2004ApJ...601..759T, 2004A&A...413..489M, 2007A&A...467..501T, 2007A&A...466..521T, 2009A&A...501..879T}, and consequently the spectrum cannot be fitted by the simple power-law model. Several authors reported that such a curved spectrum can be fitted well with a log-parabola function. 
In this work, we have used log-parabola function and energy dependent physical models as input particle density to the \emph{($synconv \otimes n(\xi)$)} model. We performed the spectral fit for each of the resultant grouped spectra obtained in the energy-range 0.3-10 keV, while we added 3\% systematics to the data in order to reduce emission model related uncertainties. During the fit, the neutral hydrogen column density, $N_{H} = 1.92\times10^{20} cm^{-2}$ was kept froze, the $N_H$ value is obtained in the LAB survey \citep{2005A&A...440..775K}.

\subsection{Log-parabola model}
Firstly, we consider the case when the underlying particle density is described by the log-parabola function, and is defined by, \\
\begin{equation}
 \label{eq:logpar}
    n(\gamma) d{\gamma} = {K} ({{\gamma}/{\gamma_{r}}})^{-{\alpha}-{\beta}{\log({{\gamma}/{\gamma_{r}}})}}d\gamma\\
\end{equation}
Here, ${\alpha}$ is the particle spectral index at the reference energy, $E_r={\gamma_{r}mc^2}$,  ${\beta}$ and ${K}$ are the curvature parameter and the normalization, respectively. The synchrotron convoluted equation involves $\xi$ parameter instead of $\gamma$, therefore, replacing $\gamma$ by $\xi/\sqrt{\mathbb{C}}$, the log-parabola function takes the form as\\
\begin{equation}
 \label{eq:logpar_conv}
    n(\xi)  = K ({{\xi}/{\xi_{r}}})^{-{\alpha}-{\beta}{\log({{\xi}/{\xi_{r}}})}}\\
\end{equation}
During the spectral fit with \emph{$synconv \otimes n(\xi)$} model (Equation \ref{eq:obs_flux}), ${\xi_r}^{2}$ was fixed at 1 keV, therefore the spectrum is determined by the three free parameters, viz. ${\alpha}$, ${\beta}$, and norm $\mathbb{N}$. From Equations  \ref{eq:obs_flux} and \ref{eq:logpar_conv},  $\mathbb{N}$  can be obtained as
\begin{equation}
 \label{eq:norm_lp}
   \mathbb{N}  = \frac{\delta^3(1+z)}{d_L^2} V \,\mathbb{A}K
\end{equation}

\begin{figure*}
\includegraphics[scale=0.5, angle=0]{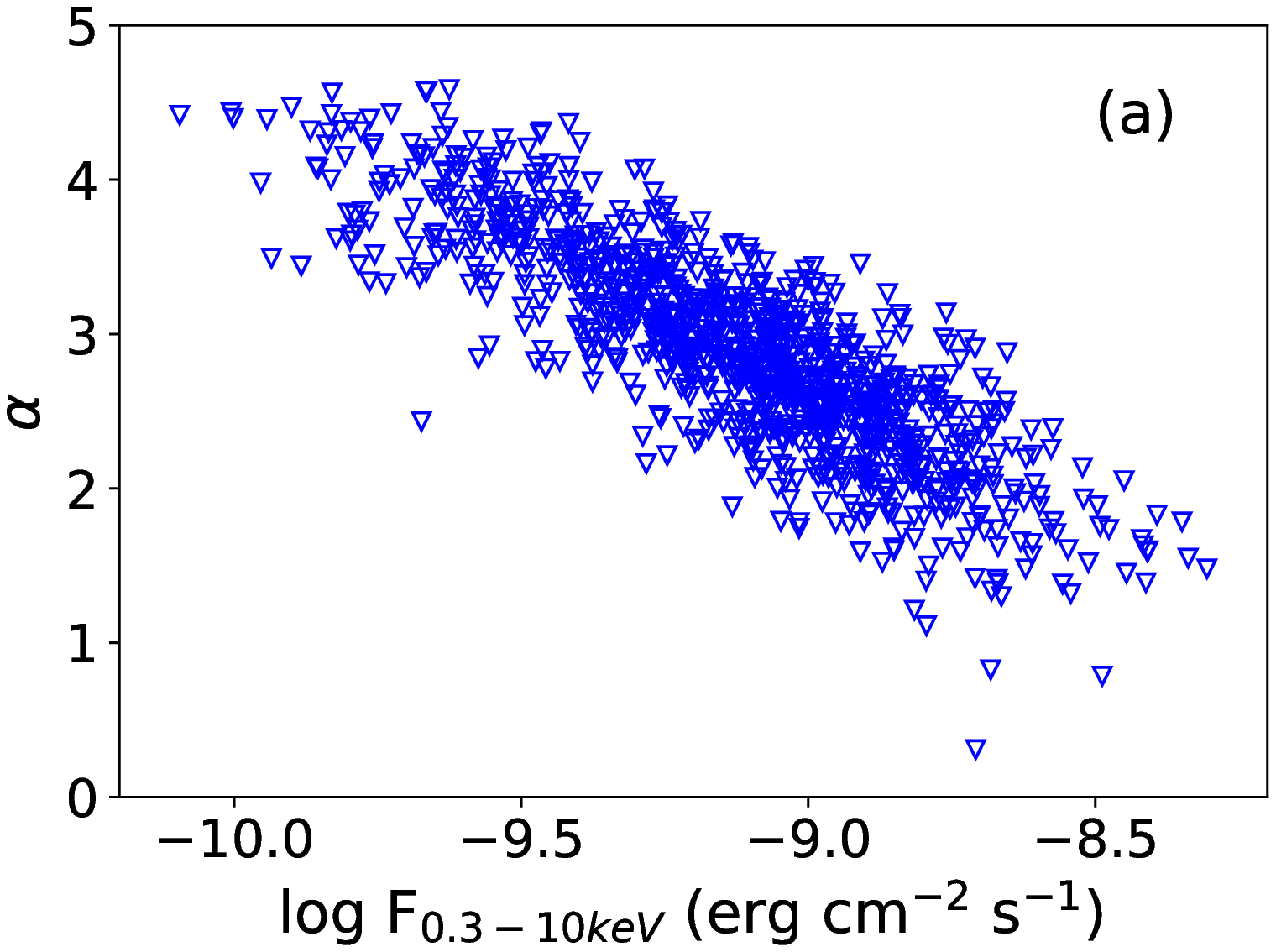}
\includegraphics[scale=0.5, angle=0]{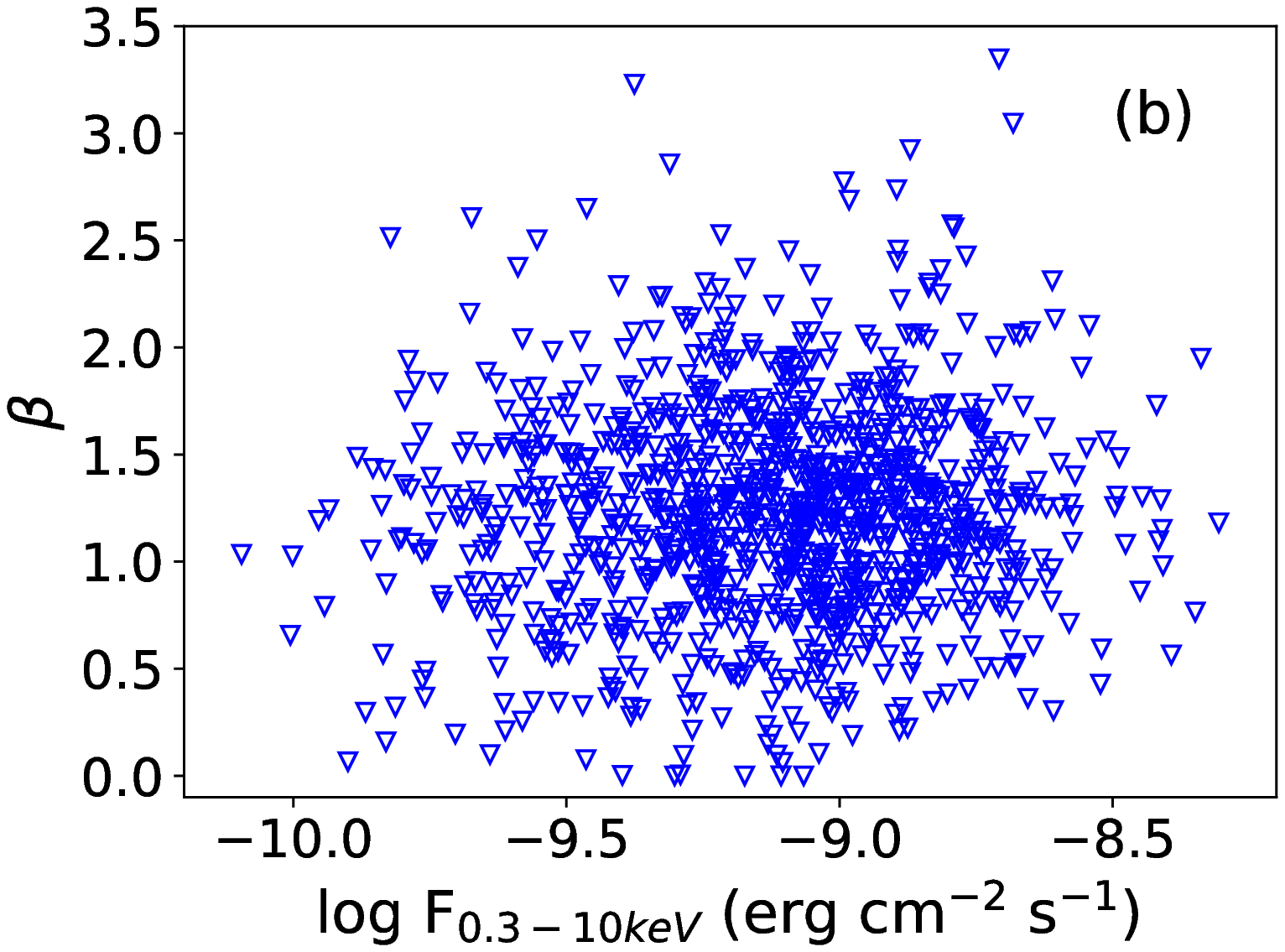}
\includegraphics[scale=0.5, angle=0]{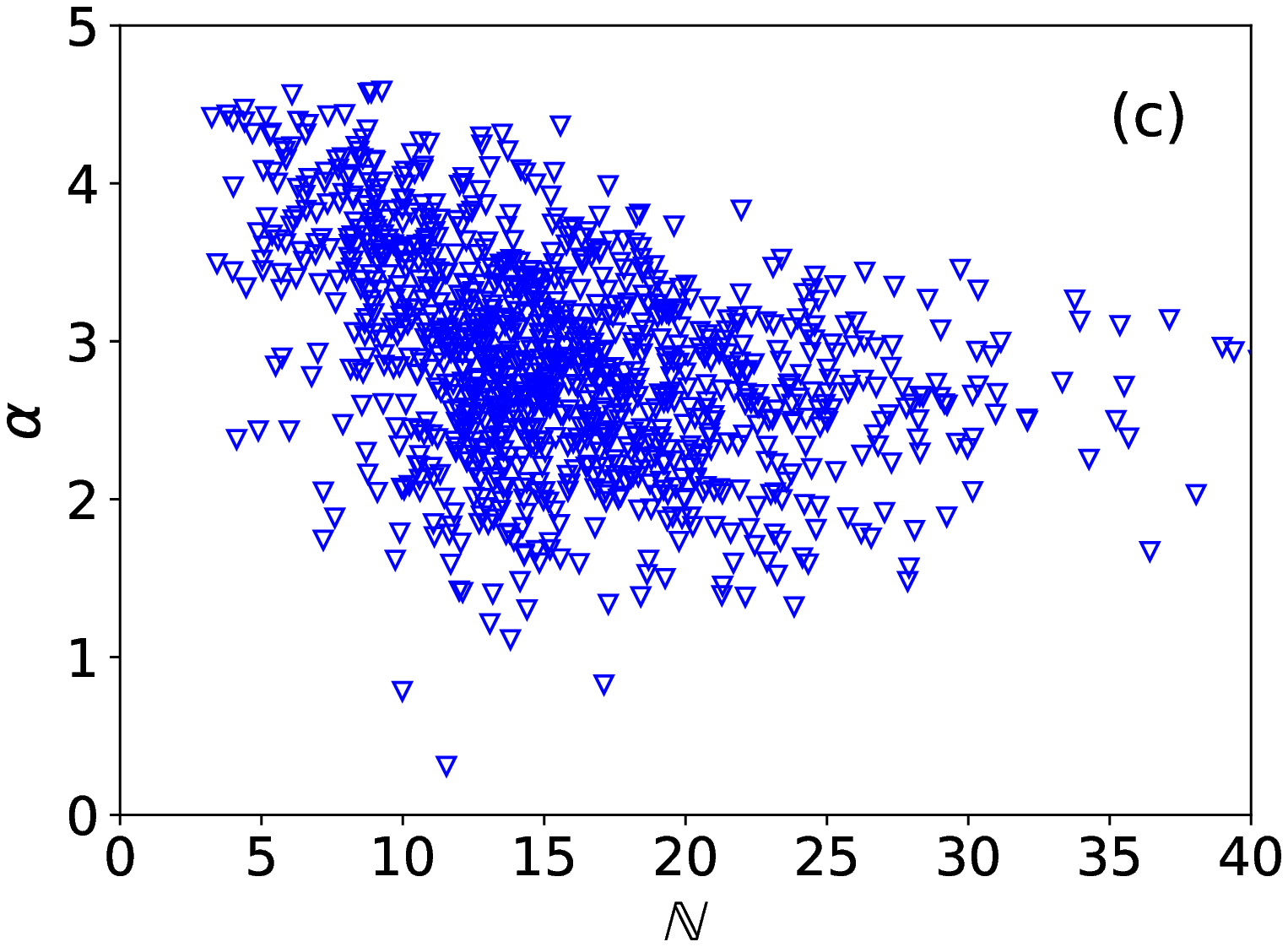}
\includegraphics[scale=0.5, angle=0]{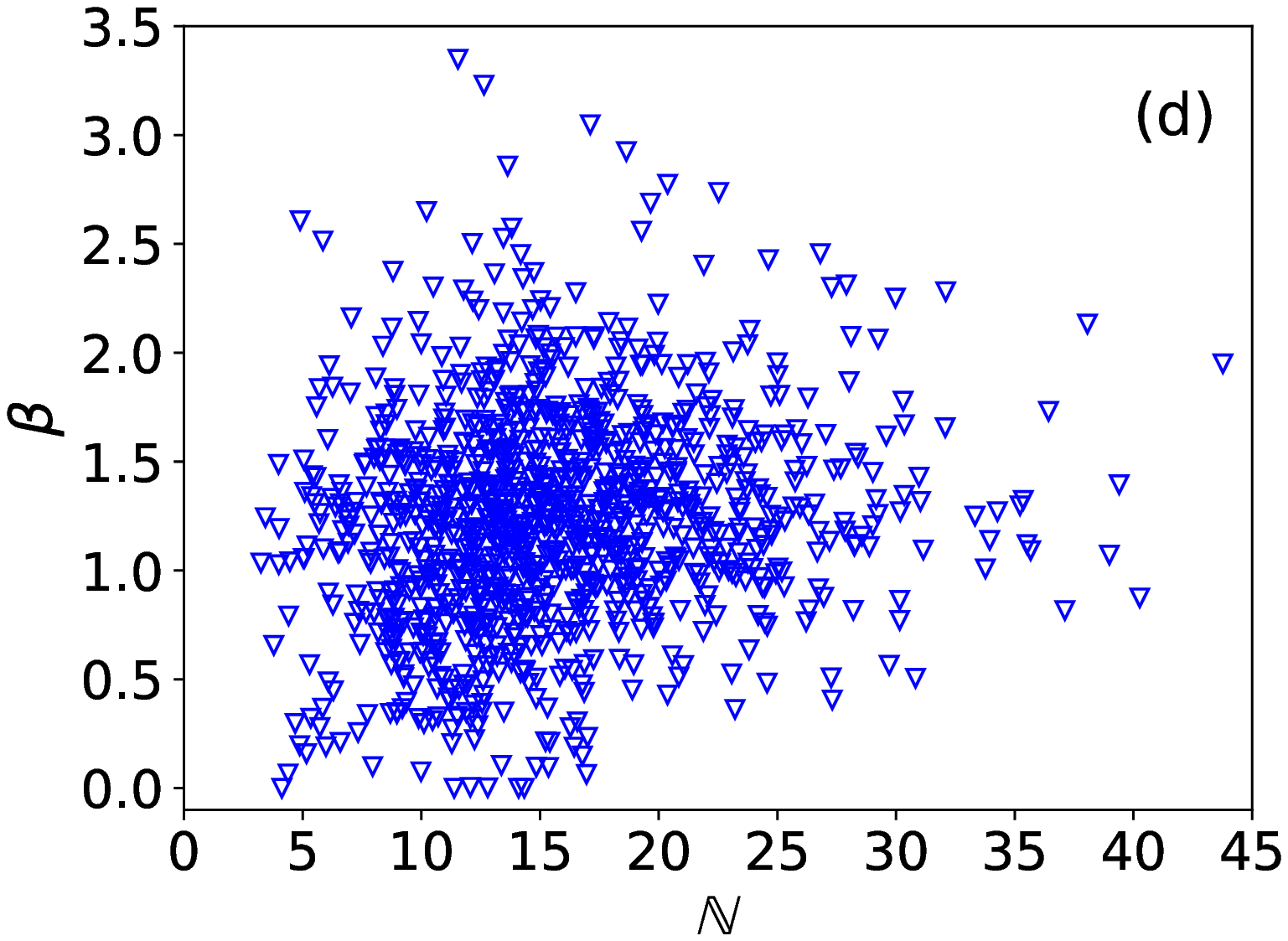}
\includegraphics[scale=0.5, angle=0]{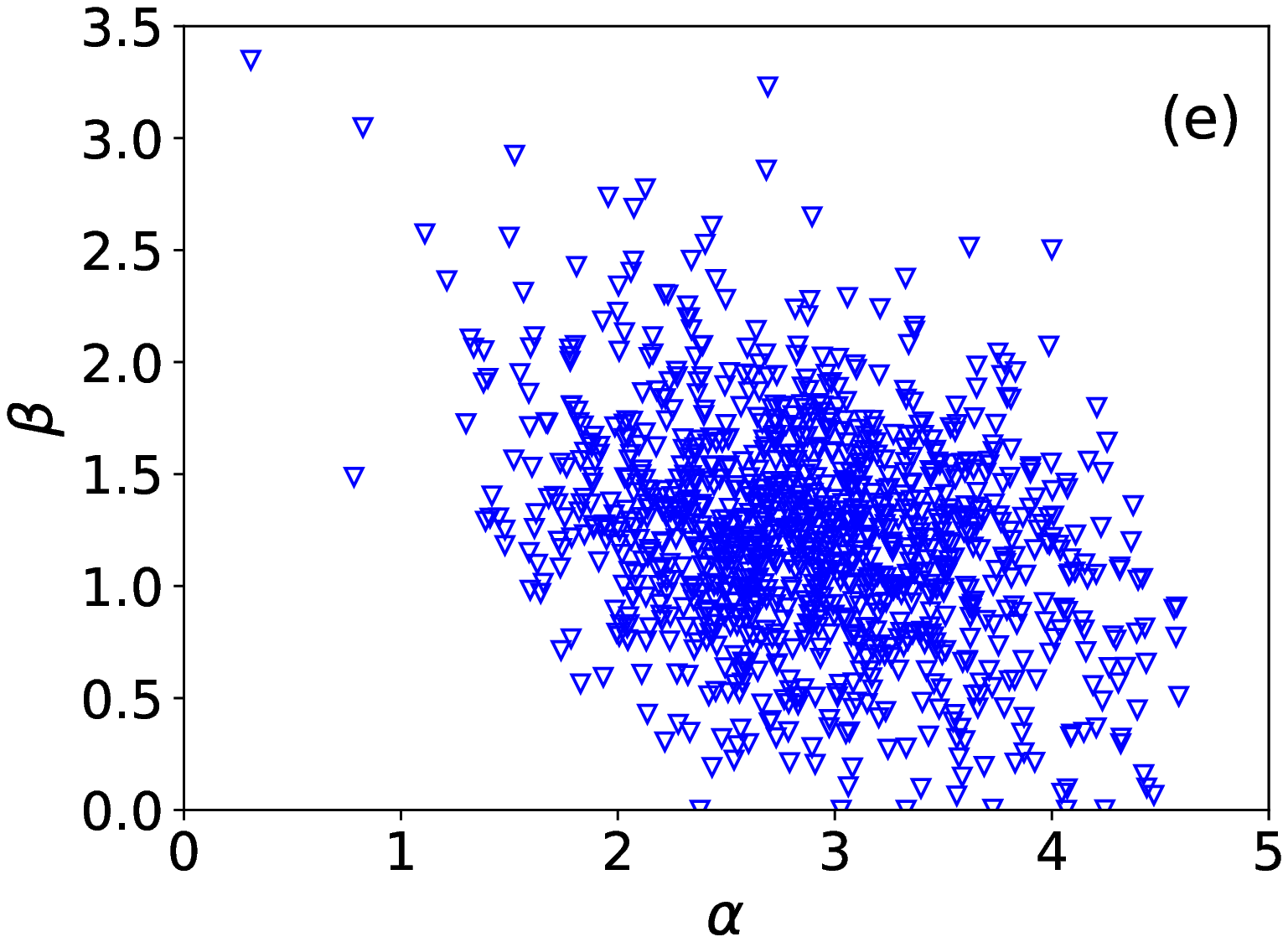}
\caption{Cross plots of the log-parabola model fit parameters and $F_{0.3-10keV}$. Panels (a-b): $\alpha$ (spectral index) and  $\beta$ (curvature parameter) are plotted vs  $F_{0.3-10keV}$. Panels (c-d): $\alpha$ and $\beta$ are plotted vs normalisation parameter. Panel: $\beta$ is plotted against $\alpha$.}

\label{fig:log}  
\end{figure*}

The obtained model is then fitted to each spectrum of the XRT observations. To assess the significance of any correlation or anticorrelation between the best fit parameters, we used Monte Carlo simulation technique. For each data point and its corresponding error in the time series, we simulated 10,000 random datasets by considering the underlying normal distribution of observed data points. The Spearman's rank correlation coefficient r$_s$ and null hypothesis probability P$_s$ were calculated using the simulated datasets. The correlation results between best fit parameters and flux are shown in Table \ref{tab:R}, where the top panel corresponds to the log-parabola model. The correlation plots between the log-parabola model fit parameters and the 0.3-10 keV flux ($F_{0.3-10keV}$), are shown in Fig. \ref{fig:log}. Analogous to the previous results, a strong anticorrelation between $\alpha$ and $F_{0.3-10keV}$ with $r_{s}$ ($P_{s}$) as $-0.81\pm0.005$ ($4.67\times{10^{-229}}$) is observed, which implies a harder when brighter behavior in the spectrum \citep{2004A&A...413..489M,2008A&A...478..395M, 2015A&A...580A.100S,2018MNRAS.480.2046G}. However, there is no correlation between $\beta$ and $F_{0.3-10keV}$ with $r_{s}$ ($P_{s}$) as $0.04\pm0.02$ (0.27). In addition, a weak anticorrelation is observed between $\alpha$ and $\beta$ with $r_{s}$ ($P_{s}$) as $-0.21\pm0.02$ ($7.68\times{10^{-9}}$), such negative trend between $\alpha$ and $\beta$ was seen by \citealp{2018MNRAS.480.2046G} and also by \citealp{2020ApJS..247...27K} in some short-term flares. On the other hand, no correlation was seen between $\alpha$ and $\beta$ during January 2013 -- June 2014 \citep{2015A&A...580A.100S, Kapanadze_2016, 2017ApJ...848..103K}. Furthermore, we obtained a nearly moderate anticorrelation between $\alpha$ and $\mathbb{N}$ with $r_{s}$ ($P_{s}$) as $-0.41\pm0.008$ ($1.78\times{10^{-39}}$), and a mild positive correlation between $\beta$ and $\mathbb{N}$ with $r_{s}$ ($P_{s}$) as  $0.15\pm0.02$ ($2.37\times{10^{-5}}$). The log-parabola model is the limited version of more general physical model under some specific scenarios, as described in sections 3.3.1 \& 3.3.2.
%However, the relation of $\alpha$ and $\beta$ with the physical parameters is not known, therefore it is difficult to obtain a physical picture which can be acquired from the correlation.  Therefore, it is important to consider the physical models capable of reproducing the curvature such that the parameter correlation will help in obtaining the physical information of the source. 
In the following subsection, we will discuss the physical models and their capability of reproducing the observed spectrum.

\subsection{Power-law particle distribution with maximum electron energy}
\begin{figure*}
\includegraphics[scale=0.5, angle=0]{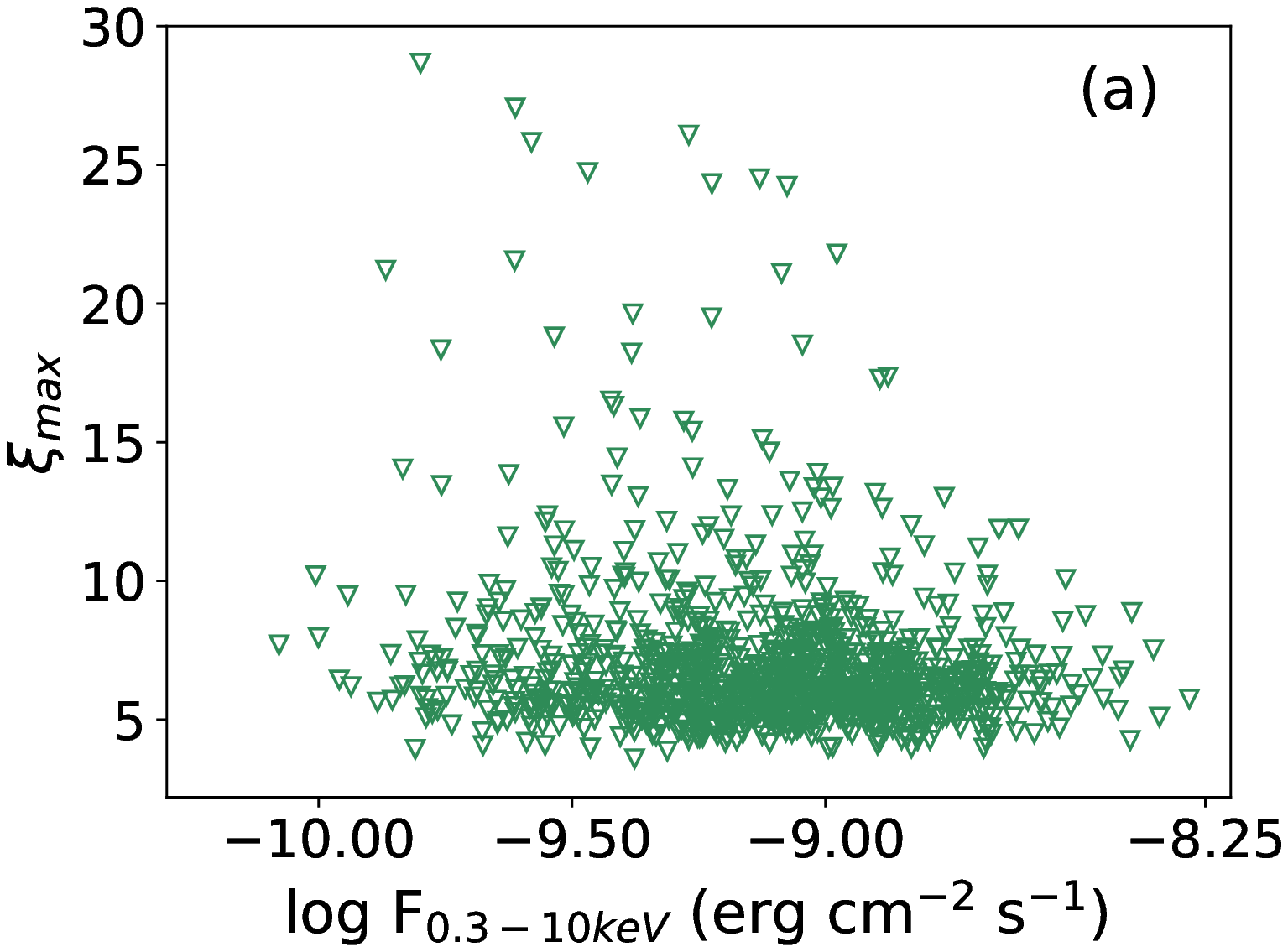}
\includegraphics[scale=0.5, angle=0]{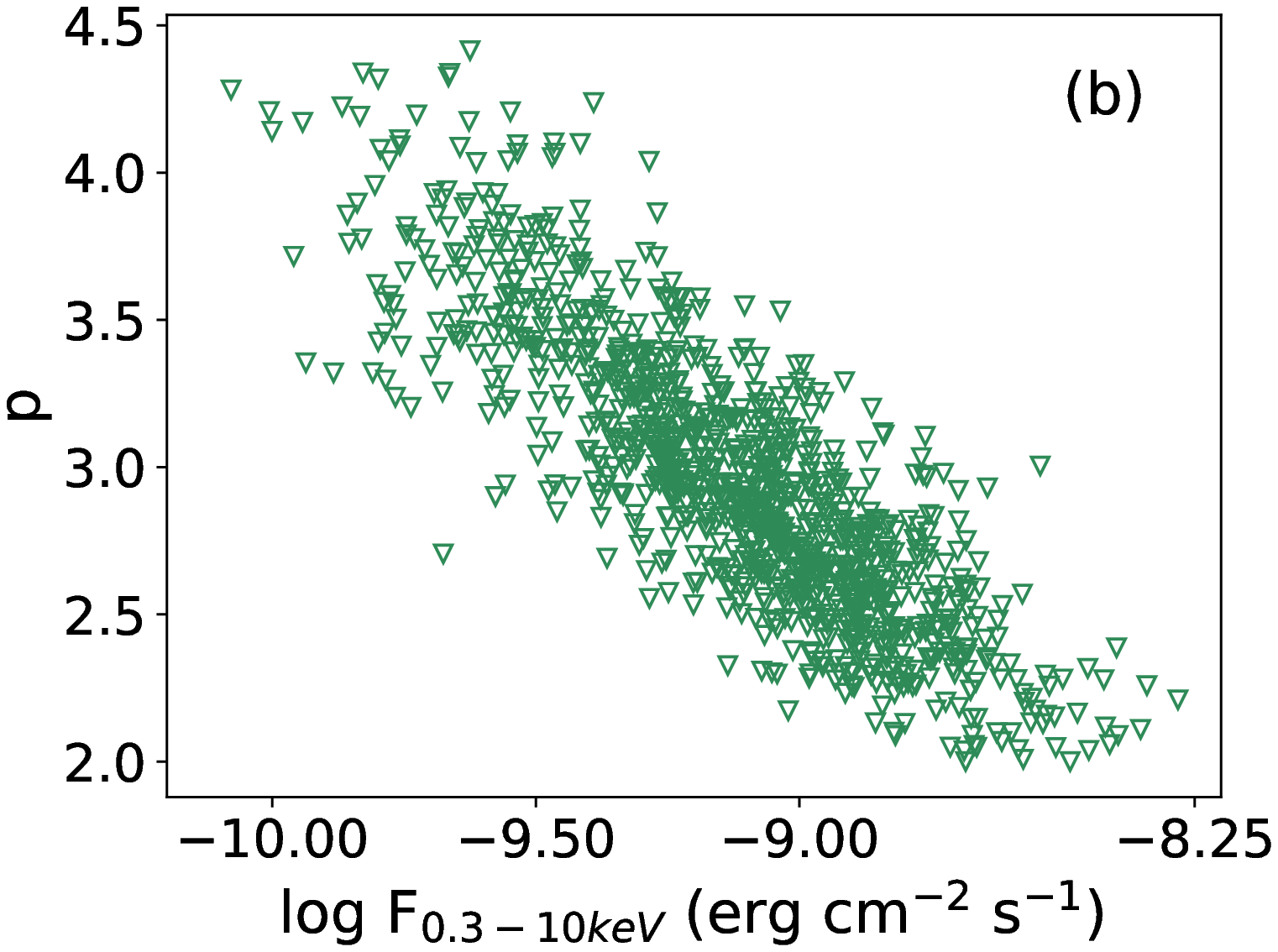}
\includegraphics[scale=0.5, angle=0]{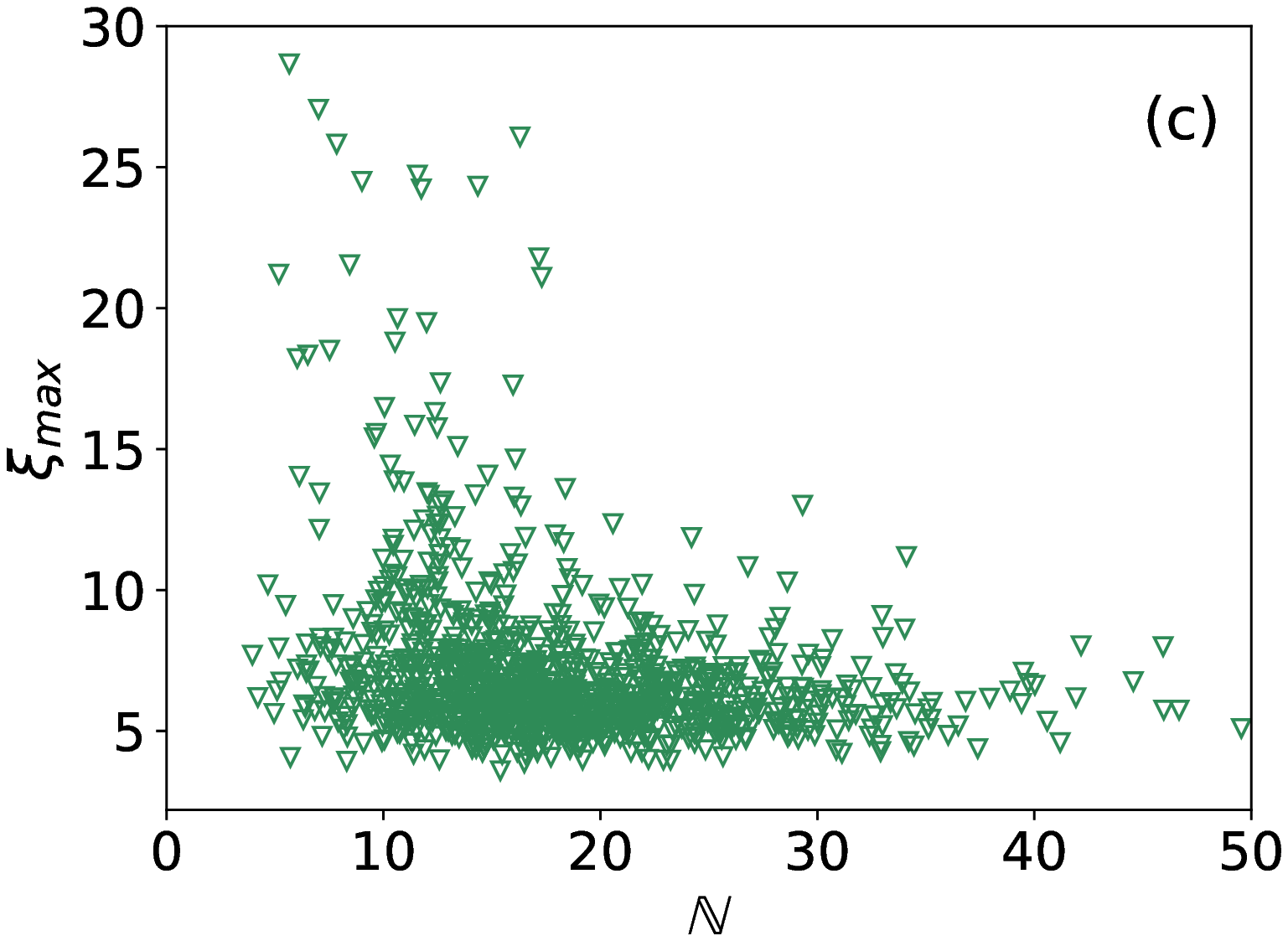}
\includegraphics[scale=0.5, angle=0]{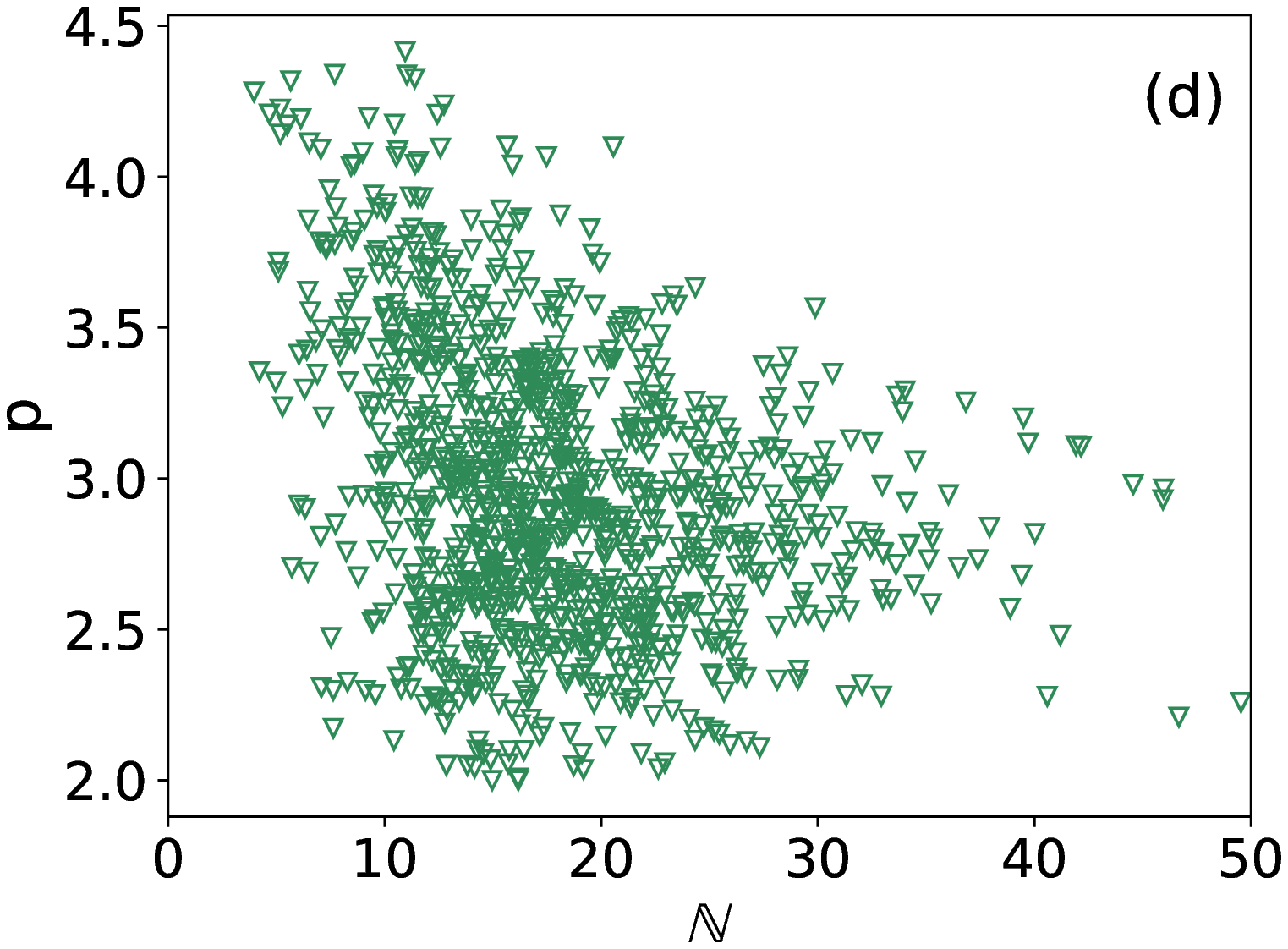}
\includegraphics[scale=0.5, angle=0]{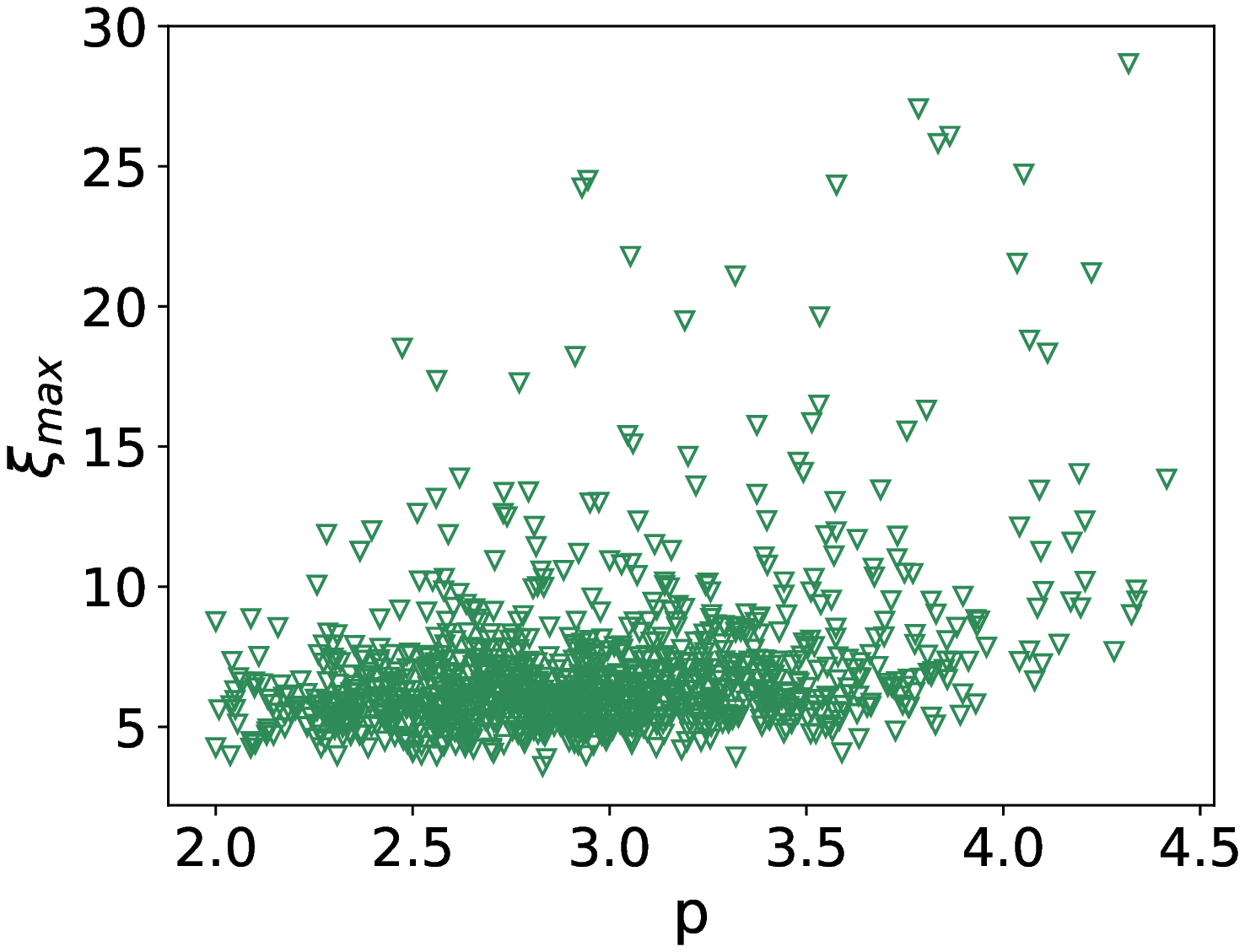}
\caption{Scatter plots between the power-law with $\xi_{max}$ model fit parameters. Panels (a-b): the maximum energy of electron ($\xi_{max}$) and the particle spectral index (p) are plotted vs flux, F$_{0.3-10 keV}$. Panel (c-d): $\xi_{max}$ and p are plotted vs normalisation. Panel e: $\xi_{max}$ is plotted vs p.} 

\label{fig:gmax} 
\end{figure*}

We refit the spectra by considering the shape of particle density as power-law model with maximum electron energy. In this case, we consider the particle acceleration mechanism and radiative losses. The  spectral curvature in the power-law particle distribution is obtained due to fast decay of emitting particles near the maximum available particle energy $\gamma_{max} mc^2$, where $\gamma_{max}$ is the maximum Lorentz factor that an electron can attain before it loses energy. Here we consider that the particles are accelerated through Fermi acceleration process near the shock front and lose energy by emitting radiation through the synchrotron process. 
%In this regard, we consider two sub-regions in the emission zone, one around the shock front where the acceleration of a mono energetic electron distributions take place and the next down stream where they lose most of their energy through radiative processes. We label the former region as acceleration region(AR) and the latter as cooling region(CR). Mono energetic electrons with Lorentz factor ${{\gamma}_{0}}$ are injected into the acceleration region, get accelerated and escape into the downstream region.
The steady-state evolution of electrons in such a region is governed by \citep{1962SvA.....6..317K},
\begin{equation}
	\label{eq:arkinetic}
	{\frac{\partial}{\partial\gamma}} \left[\left(\frac{\gamma}{t_{acc}}-\Lambda_a\gamma^2\right)n_{a}\right]+\frac{n_{a}}{t_{esc}}= Q_o\delta(\gamma-\gamma_0)  
\end{equation}
%\begin{equation}
%	{\rm CR:} \frac{d}{d\gamma} [B{\gamma^2}n_{s}(\gamma)]+\frac{n(\gamma)}{t_{esc}}=\frac{n_{s}(\gamma)}{t_{esc}}
%	\label{eq:crkinetic}
%\end{equation}
where $\Lambda_a=\frac{1}{{\gamma_{max}}{t_{acc}}}$, 
${t_{acc}}$ and ${t_{esc}}$ are the acceleration and escape time scales of electrons, 
 $\Lambda_a\gamma^2$  describes the radiative energy loss rate. If we consider the case of ${t_{acc}}$ and ${t_{esc}}$ being energy independent, such that their ratio is defined as,
\begin{equation}
	\frac{t_{acc}}{t_{esc}} = p - 1
\end{equation}
where p is the particle spectral index, then the steady-state solution of Equation \ref{eq:arkinetic} can be obtained as \citep{kirk}, \\
\begin{equation}\label{eq:gmaxsoln}
n(\gamma) d\gamma = K \gamma^{-p} \left(1-\frac{\gamma}{\gamma_{max}}\right)^{(p-2)} d\gamma
\end{equation}

where $K=Q_0t_a\gamma_0^{p-1}\mathbb{C}^{p/2}$, here $Q_0$ is mono-energetic injection
at minimum energy, $\gamma_0$. 
Again the \emph{$synconv \otimes n(\xi)$} model (Equation \ref{eq:obs_flux}) used for spectral fit contains $\xi$ instead of $\gamma$, therefore replacing $\gamma$ by $\xi/\sqrt{C}$ in \ref{eq:gmaxsoln}, we have 
\begin{equation}\label{eq:gmax_part}
n(\xi) = K \xi^{-p} \left(1-\frac{\xi}{\xi_{max}}\right)^{(p-2)} 
\end{equation}
where  $\xi_{max}=\gamma_{max}\sqrt{\mathbb{C}}$. We performed the fit with the convolved spectral model, \emph{$synconv \otimes n(\xi)$}; where $n(\xi)$ is given by Equation \ref{eq:gmax_part} (hereafter $\xi_{max}$ model), for all the available \emph{Swift}-XRT observations. The fit is carried with three parameters viz. norm $\mathbb{N}$, $\xi_{max}$ and p. In this model, the number of free parameters are same as the log-parabola model. Here $\mathbb{N}$ is defined as
\begin{equation}
 \label{eq:norm_gmax}
   \mathbb{N}  = \frac{\delta^3(1+z)}{d_L^2} V \,\mathbb{A}Q_0t_{acc}\gamma_0^{p-1}\mathbb{C}^{p/2}
\end{equation}
The model provides a reasonable fit to the X-ray spectrum in the energy range 0.3--10 keV, and the reduced-$\chi^2$ values obtained from the $\xi_{max}$ model are equally good as those obtained from the log-parabola model. The difference of the reduced-$\chi^{2}$ values for the log-parabola and the $\xi_{max}$ model, vs. reduced-$\chi^{2}$ of the log-parabola model is shown in Fig. \ref{fig:redchi}(a).  
The correlation results between the $\xi_{max}$ model parameters are reported in the second rows of Table \ref{tab:R}, and the plots are shown in Fig. \ref{fig:gmax}. A strong anticorrelation is observed between p and flux, F$_{0.3-10 keV}$ with 
$r_{s}$ ($P_{s}$) as $-0.80\pm0.007$ ($1.92\times{10^{-209}}$), while no correlation is obtained between $\xi_{max}$ and F$_{0.3-10 keV}$ with $r_{s}$ ($P_{s}$) as $-0.03\pm0.02$ (0.44). A weak anticorrelation is obtained between p and $\mathbb{N}$ with $r_{s}$ ($P_{s}$) as $-0.28\pm0.009$ ($6.31\times{10^{-19}}$), and $\xi_{max}$ vs. $\mathbb{N}$ with $r_{s}$ ($P_{s}$) as $-0.15\pm0.02$ ($2.28\times{10^{-4}}$). The weak correlation between  $\xi_{max}$ and $\mathbb{N}$ can be expected as the functional form of $\mathbb{N}$ (Equation \ref{eq:norm_gmax}) is independent of $\xi_{max}$. However, the weak anticorrelation obtained between p and $\mathbb{N}$ contradicts with the functional form of $\mathbb{N}$ (as $\log\mathbb{N}\propto p$ see Equation \ref{eq:norm_gmax}). 
Moreover, a weak positive correlation is observed between $\xi_{max}$ and p with $r_{s}$ ($P_{s}$) as $0.20\pm0.02$ ($4.94\times{10^{-7}}$). In the Fermi acceleration scenario, the $\xi_{max}$ is decided by the acceleration rate and radiative loss such that
$\xi_{max} \propto \frac{1}{t_{acc} B^{3/2}}$ and $p \propto {t_{acc}}$. Therefore, an anticorrelation is expected between $\xi_{max}$ and p, which disputes with the weak positive correlation observed between $\xi_{max}$ and p. These disputes suggest that the PL with maximum electron energy model is not suitable for reproducing the observed synchrotron spectrum. However, the model can be consistent with the observations if we consider the following conditions. The acceleration time-scale varies with the magnetic field such as $t_{acc}\propto {B^{-n}}$ or $B \propto (p-1)^{-\frac{1}{n}}$. Consequently, $\xi_{max} \propto B^{n-3/2} \propto (p-1)^{-(n-3/2)/n}$. %if the $t_{acc}$ is independent of magnetic field, B then 
Therefore, a positive correlation is expected between $\xi_{max}$ and p if $n < \frac{3}{2}$ and linear correlation is expected if $n\sim \frac{3}{4}$. 

 \subsection{Energy dependent model:}

Alternatively, the physical models with the energy dependent escape time-scale [$t_{esc}(\gamma)$] or the energy dependent acceleration time-scale [$t_{acc}(\gamma)$], can also explain the spectral curvature. Again we consider that the electrons gain energy mainly by crossing the shock front such that $t_{acc}$ is determined by the time-scale at which particles cycle across the shock and these accelerated electrons diffuse away from shock region at a rate $1/t_{esc}$, move to the downstream region and finally lose energy by emitting the synchrotron radiations.
%Now we have two scenarios viz. energy dependent escape and energy dependent acceleration:
\subsubsection{$t_{esc}$ is energy dependent (EDD model)}
\begin{figure*}
\includegraphics[scale=0.5, angle=0]{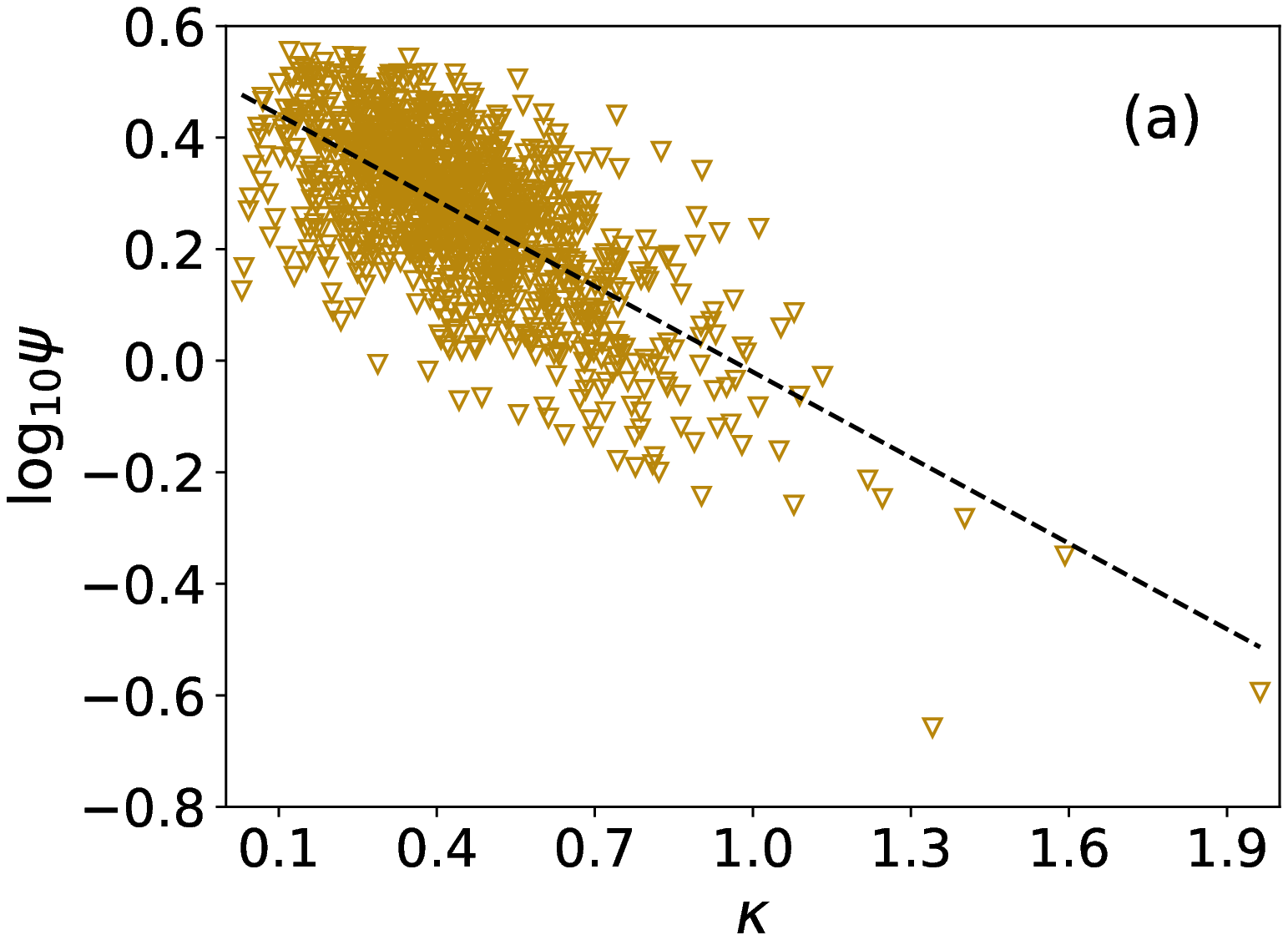}
\includegraphics[scale=0.5, angle=0]{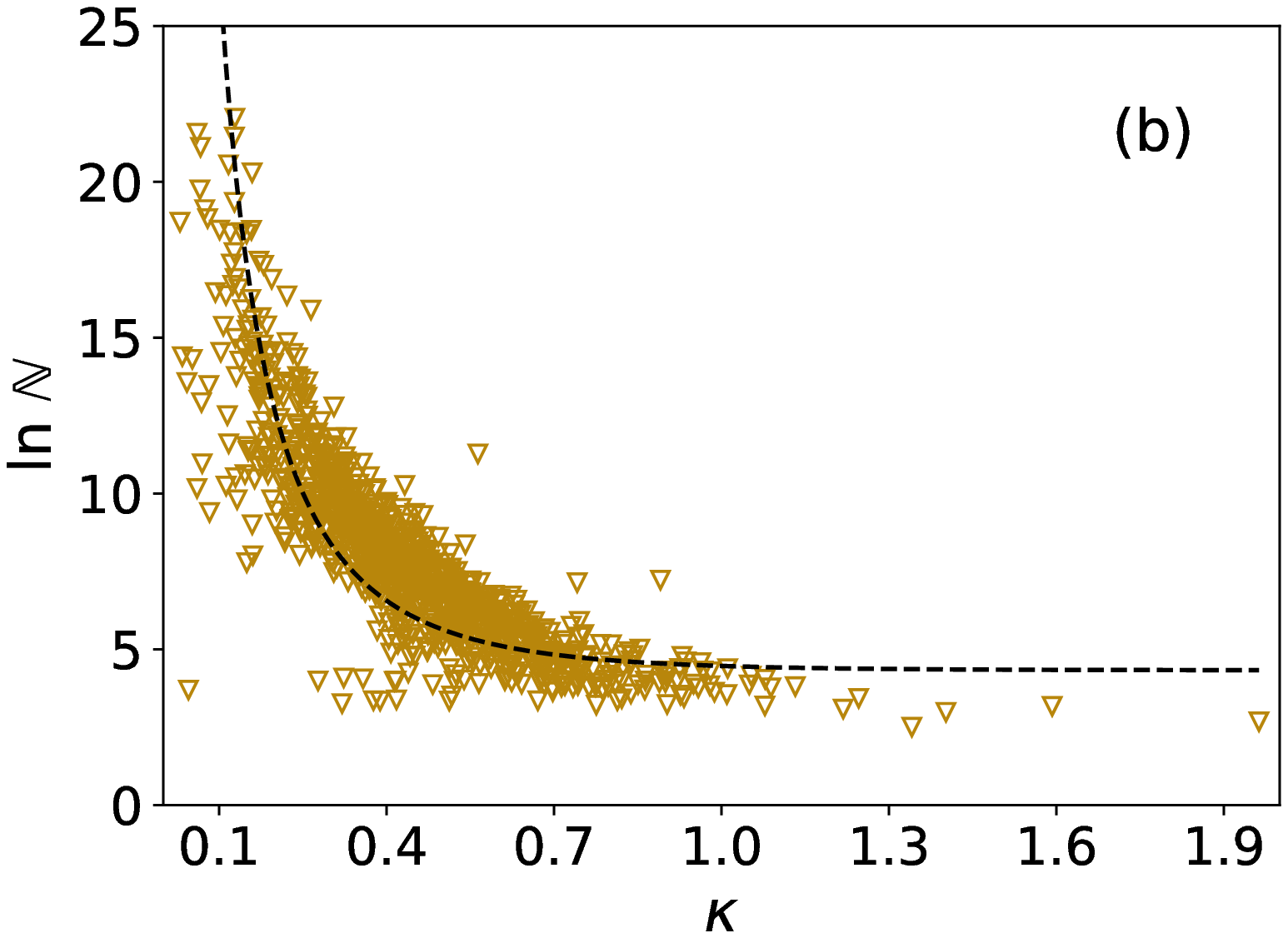}
\includegraphics[scale=0.5, angle=0]{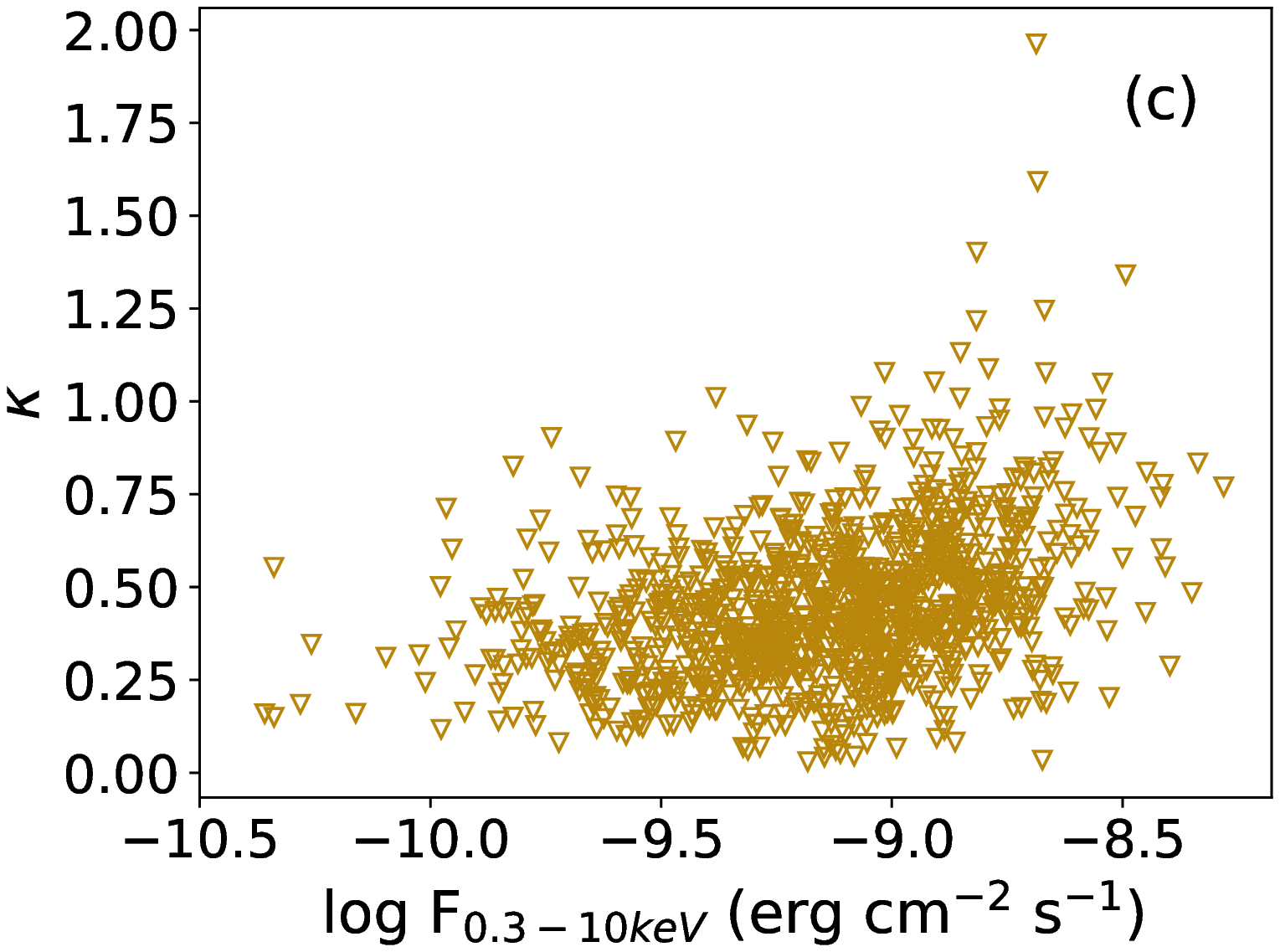}
\includegraphics[scale=0.5, angle=0]{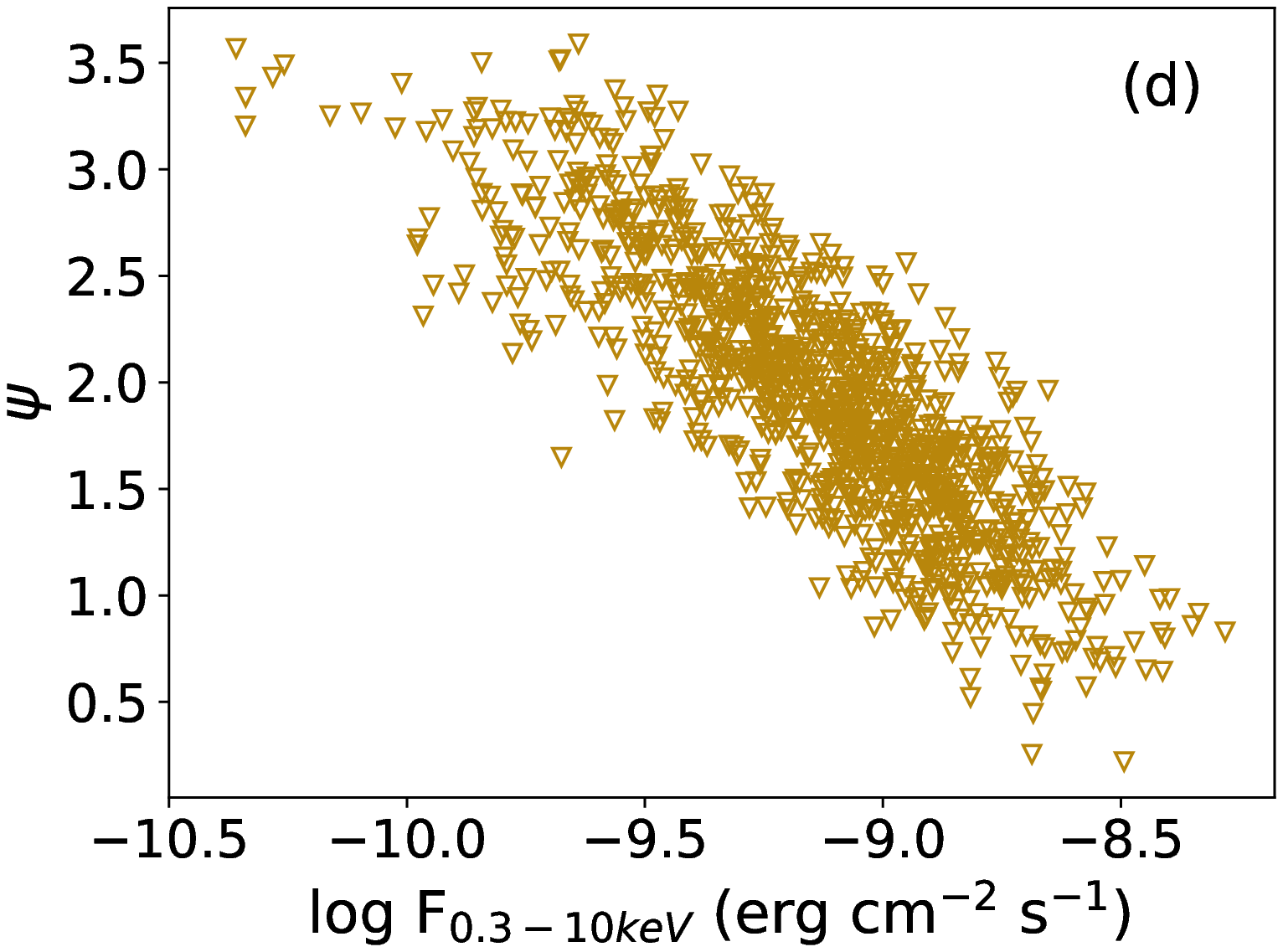}

\caption{Scatter plots between the energy-dependent $t_{esc}$ model parameters. Panels (a-b): log$_{10}{\psi}$ and log$\mathbb{N}$ are plotted vs ${\kappa}$. Panel (c-d): $\kappa$ and $\psi$ are plotted against flux, F$_{0.3-10 keV}$. }
%Panel e: ${\kappa}$ is plotted against ${\psi}$. } 

\label{fig:edd} 
\end{figure*}

%First we consider the case  where the magnetohydrodynamic turbulence in jet flow causes the spatial diffusion of the electron, $t_{esc}$ to be energy-dependent while ${t_{acc}}$ energy independent, such that the quantity $t_{esc}$ is defined as\\
In the jet environment, the diffusion occurs in the region filled with magnetic field, which can make the escape time scale dependent on the gyration radius of electron. This inturn can make escape time scale energy dependent. Here we parameterize the energy-dependent as\\

\begin{equation}\label{eq:energy_dep}
	{t_{esc}} = {t_{esc,R}}\left(\frac{\gamma}{\gamma_R}\right)^{-\kappa}
\end{equation}
where, ${t_{esc,R}}$ corresponds to ${t_{esc}}$ when the  electron energy is $\gamma_R mc^2$, and $\kappa$ decides on the energy dependence of the escape of electrons. Here, ${t_{esc}}$ can not be larger than the free streaming value, ${t_{esc,R}}$, and this set limits on $\gamma$ to  $\gamma < \gamma_{R}$ . Consequently for $\gamma_{0}<\gamma$ and if we neglect the synchrotron energy loss, the electron energy distribution with the energy dependence of ${t_{esc}}$ will take the form \citep{2021MNRAS.tmp.2632H},\\

\begin{equation}\label{eq:eddcr}
  n(\xi)=Q_o t_{acc}\sqrt{\mathbb{C}} \xi^{-1}\rm{exp}\left[-\frac{\eta_R}{\kappa}\,\left(\left(\frac{\xi}{\xi_R}\right)^{\kappa}-\left(\frac{\xi_0}{\xi_R}\right)^{\kappa}\right)\right]
\end{equation}
  %\label{n_EDDt}
%\begin{equation}\label{eq:eddcr}
%	n(\xi)=K \mathbb{C} \xi^{-2}\rm{exp}\left[-\frac{\psi}{\kappa}\,\xi^{\kappa}\right]%\quad\rm{for}\quad \gamma\gg\frac{1}{Bt_{esc}}
%\end{equation}
Here, $\eta_{R}\equiv {t_{acc}}/{t_{esc,R}}$, $\xi_{R}={\sqrt{\mathbb{C}}}\gamma_{R}$ and $\xi_{0}={\sqrt{\mathbb{C}}}\gamma_{0}$. In case of $\kappa << 1$, the particle distribution will be represented by a log-parabola distribution such that, $n(\xi) \propto (\xi/\xi_R)^{-\eta_R-1 -\eta_R\kappa log(\xi/\xi_R)}$, while for the case $\xi \rightarrow 0$, the solution will be identical to the one given in Equation \ref{eq:gmax_part} for $p=1+\eta_{R}$ and $\gamma << \gamma_{max}$ i.e., the case when escape time-scale will be energy independent or equal to the free streaming value, ${t_{esc,R}}$. The details of these assumptions are given in \citealp{2021MNRAS.tmp.2632H}. During the spectral fitting, there is degeneracy in the parameters associated with Equation \ref{eq:eddcr}. Therefore, we remove the degenerate parameters and use the modified Equation  as follows,

\begin{equation}\label{eq:eddcr1}
	n(\xi)=K{\sqrt{\mathbb{C}}} \xi^{-1}\rm{exp}\left[-\frac{\psi}{\kappa}\,\xi^{\kappa}\right]%\quad\rm{for}\quad \gamma\gg\frac{1}{Bt_{esc}}
\end{equation}
where 
\begin{equation}\label{eq:psi_kappa}
\psi=\eta_R\left(\frac{1}{\mathbb{C}\gamma_R^2}\right)^{\kappa/2}=\frac{\eta_R}{{\xi_{R}}^{\kappa}}
\end{equation}
and the normalization K is given below 
\begin{equation}
K = Q_0\,t_{acc}\,\rm{exp}\left[\frac{\eta_R}{\kappa}\,\left(\frac{\gamma_0}{\gamma_R}\right)^{\kappa}\right]
\end{equation}
The free parameters of \emph{$synconv \otimes n(\xi)$} model in this case are norm $\mathbb{N}$, $\psi$ and $\kappa$, where $\mathbb{N}$ is defined as
\begin{equation}\label{eq:norm_eed}
   \mathbb{N}  = \frac{\delta^3(1+z)}{d_L^2} V \mathbb{A}K{\sqrt{\mathbb{C}}}
\end{equation}
While carrying the spectral fit of the available Swift-XRT observations of Mkn\,421 with \emph{$synconv \otimes n(\xi)$} model, $n(\xi)$ is given by Equation \ref{eq:eddcr1}. In Fig. \ref{fig:redchi}(b), we have plotted the difference of the reduced-$\chi^{2}$ values for the log-parabola and the energy-dependent $t_{esc}$ model vs. reduced-$\chi^{2}$ values of the log-parabola model. The figure shows that both the log-parabola and the EDD model fit the X-ray spectrum well. 
The Spearman-rank correlation results between the EDD model fit parameters are presented in the third rows of Table \ref{tab:R}, and the correlation plots are shown in Fig. \ref{fig:edd}. A low positive correlation is observed between $\kappa$ and flux,  $F_{0.3-10 keV}$ with $r_{s}$ ($P_{s}$) and $0.32\pm0.02$ ($3.30\times{10^{-19}}$), while a strong negative correlation is observed between $\psi$ and flux, $F_{0.3-10 keV}$ with $r_{s}$ ($P_{s}$) values as $-0.76\pm0.008$ ($1.01\times{10^{-180}}$) respectively. \autoref{eq:psi_kappa} suggests an antirelation between $\log_{10}\psi$ and $\kappa$, which is seen in scatter plot between  $\log_{10}\psi$ and $\kappa$ (see Fig. \ref{fig:edd} (a)), the Spearman rank correlation method shows a moderate anticorrelation with $r_{s}$ ($P_{s}$) as $\sim$ $-0.51\pm0.01$ ($3.65\times{10^{-51}}$). To obtain the value of $\xi_R$, we  express \autoref{eq:psi_kappa} as
\begin{equation}\label{eq:log_psi_kappa}
\log_{10} \psi= \log_{10}\eta_R-\kappa\,{\log_{10}\xi_R}
\end{equation}
The $\chi^2$-fit of \autoref{eq:log_psi_kappa} to the scatter plot resulted in slope, ${\log_{10}\xi_R} = 0.52$ and the y-intercept, $\log_{10}\eta = 0.49$, which implies that ${\xi}_{R}$ ${\sim}$ 3.31 keV and $\eta_{R}$  ${\sim}$ 3.09. Therefore, the photon energy corresponding to $\gamma_{R}$ will be $\xi_R^2 \sim$   10.96 keV which is slightly higher than the energy range, 0.3--10 keV considered for the spectral fitting. Furthermore, the correlation between the best-fit parameters $\kappa$ and $\mathbb{N}$ can be used to estimate the energy of the electron at which they are injected. The normalization in Equation \ref{eq:norm_eed} can be written as 
 \begin{equation}\label{eq:log_norm_kappa}
\log \mathbb{N}= \frac{\eta_R}{\kappa}\,A^{\kappa}+B
\end{equation}
We fit the plot shown in Fig. \ref{fig:edd}(b) with the above equation which results A = 0.045, B = 4.33, and ${\eta_R}$ = 3.09, implies the value of $\gamma_0$ $\sim$ 0.045 $\gamma_R$, which is significantly smaller than $\gamma_R$.

\subsubsection{$t_{acc}$ is energy dependent (EDA model)}
\begin{figure*}
\includegraphics[scale=0.5, angle=0]{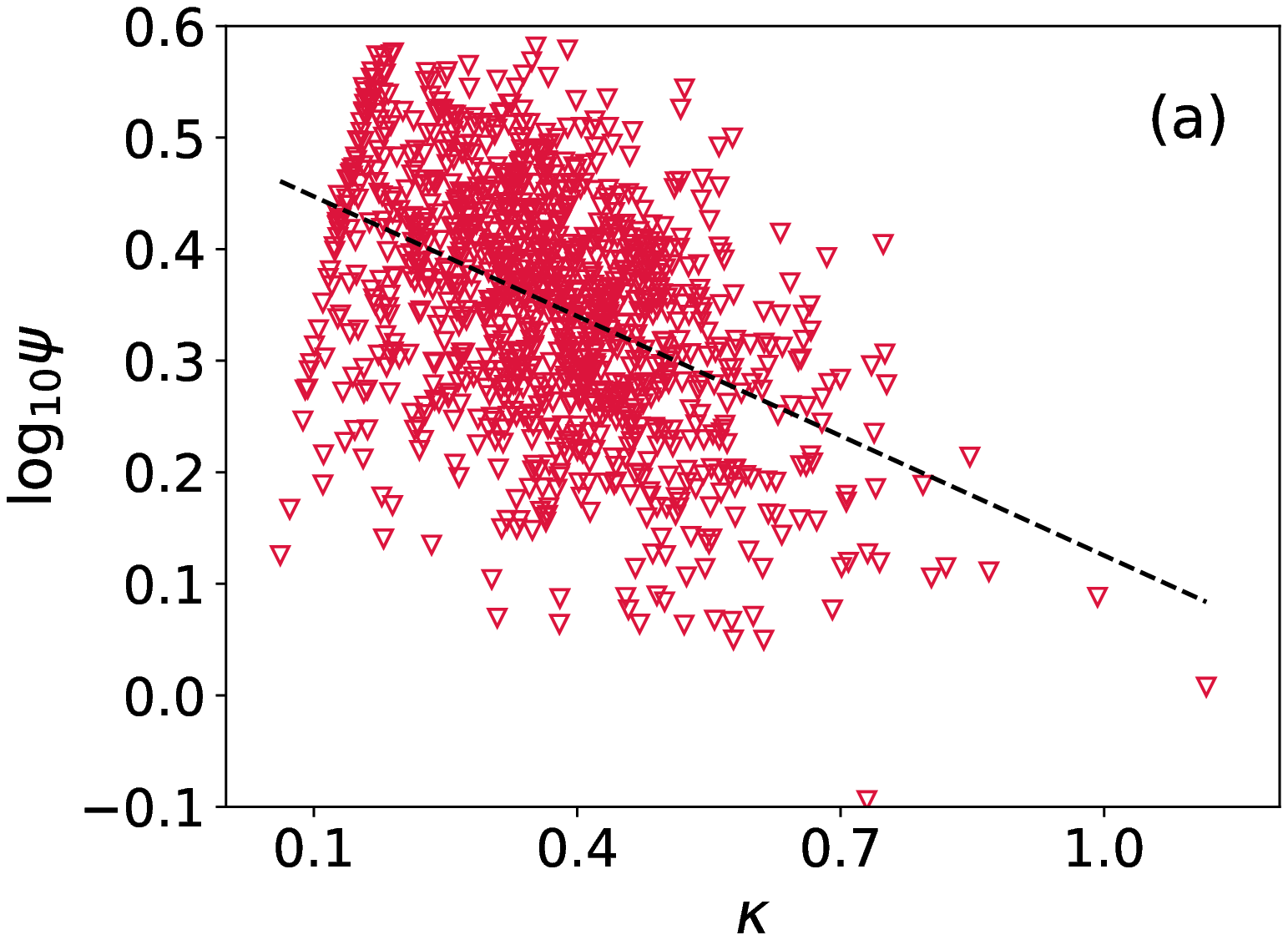}
\includegraphics[scale=0.5, angle=0]{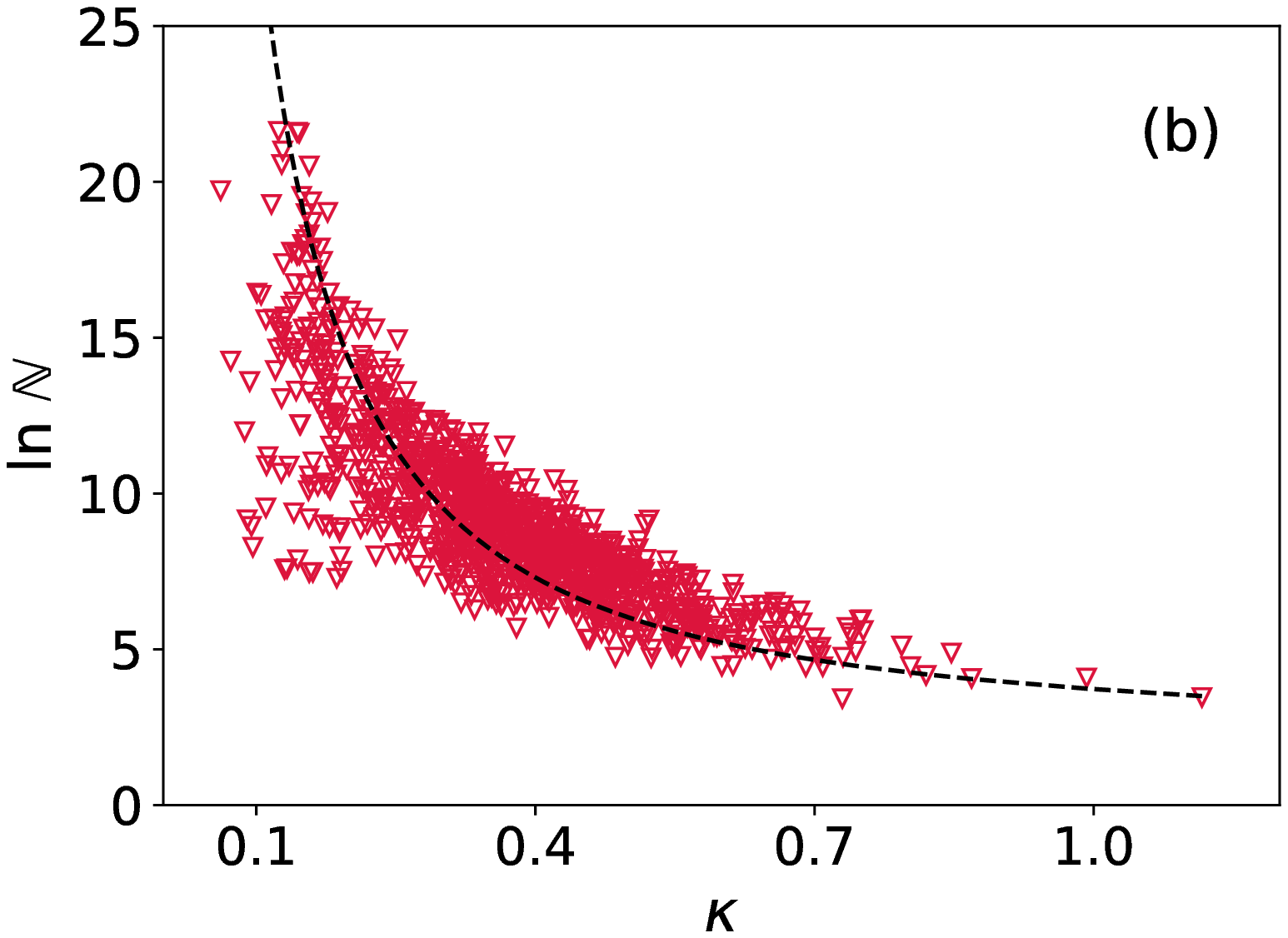}
\includegraphics[scale=0.5, angle=0]{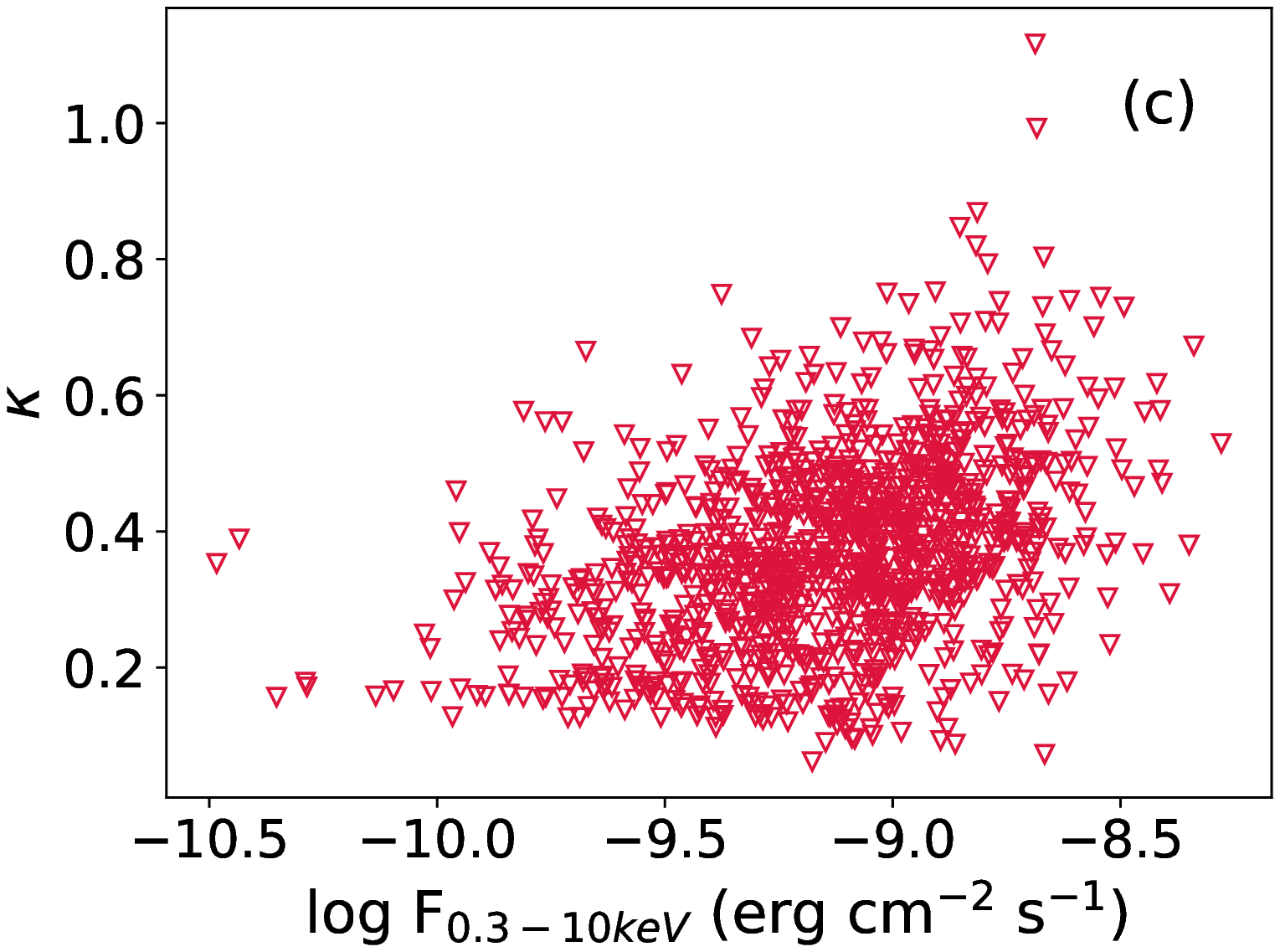}
\includegraphics[scale=0.5, angle=0]{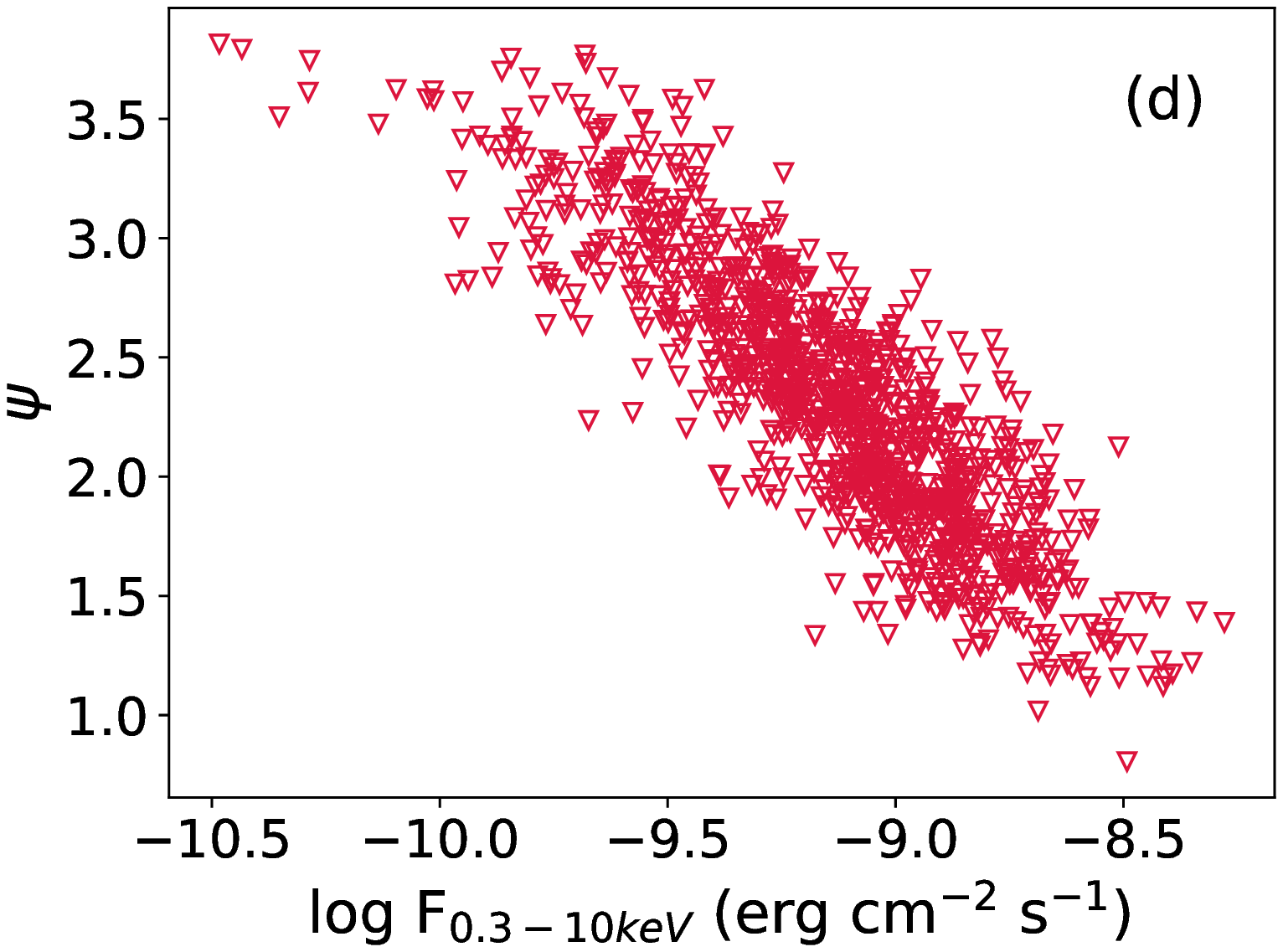}

\caption{Scatter plots between the energy-dependent $t_{acc}$ model parameters. Panels (a-b): log$_{10}{\psi}$ and log$\mathbb{N}$ are plotted vs ${\kappa}$. Panel (c-d): $\kappa$ and $\psi$ are plotted against flux, F$_{0.3-10 keV}$. }
%Panel e: ${\kappa}$ is plotted against ${\psi}$. } 

\label{fig:eda} 
\end{figure*}

Here, we consider a scenario in which radiative loss due to synchrotron emission happens in the vicinity of the shock front instead of  particles losing energy in the downstream flow as in the EDD model. In this case, we assume that the magnetohydrodynamic turbulence in jet flow makes $t_{acc}$ to be energy-dependent while ${t_{esc}}$ to be energy independent such that the energy dependence of $t_{acc}$ is defined by ${t_{acc}} = {t_{acc,R}}\left(\frac{\gamma}{\gamma_R}\right)^{\kappa}$, here $\kappa$ decides on the energy dependence of $t_{acc}$. Using Equation \ref{eq:arkinetic}, the steady-state solution of kinetic equation describing the particle distribution when $t_{acc}$ is energy dependent and ignoring the synchrotron energy loss.
%in the acceleration region for $\gamma_{0}<\gamma\ll\gamma_{max}$ can be obtained as 
\\
%	n(\xi)=K \frac{\xi^{\kappa-1}}{\mathbb{C}^{(\kappa-1)/2}} \rm{exp}\left[-\frac{\psi}{\kappa}\xi^{\kappa}\right]
\begin{equation}\label{eq:edacr}
n(\xi)=Q_0t_{acc,R}\sqrt{\mathbb{C}}\xi_R^{-\kappa}\xi^{\kappa-1}\exp \left[-\frac{\eta_R}{\kappa}\left(\left(\frac{\xi}{\xi_R}\right)^\kappa-\left(\frac{\xi_0}{\xi_R}\right)^\kappa\right)\right]
\end{equation}
%\\
%\begin{equation}\label{eq:edatg}
%	n(\gamma)=K \gamma^{\kappa-1} \rm{exp}\left[-\frac{\eta_i}{\kappa}\,\left(\frac{\gamma}{\gamma_i}\right)^{\kappa}\right]
%\end{equation}
Here, $\eta_{R}\equiv {t_{acc,R}}/{t_{esc}}$, $\xi_{R}={\sqrt{\mathbb{C}}}\gamma_{R}$ and $\xi_{0}={\sqrt{\mathbb{C}}}\gamma_{0}$. For $\kappa << 1$, the particle distribution will again represent a log-parabola distribution such that, $n(\xi) \propto (\xi/\xi_R)^{-\eta_R+\kappa-1 -\eta_R\kappa log(\xi/\xi_R)}$, and for the case $\kappa \rightarrow 0$ and $\gamma << \gamma_{max}$, the solution is similar to Equation \ref{eq:gmax_part}. 
%Again to remove the degeneracy associated with the parameters in Equation \ref{eq:edatxi}. Therefore, we remove the denegerate parameters and use the modify Equation  as follows,
The particle energy distribution will reduce to the below form, when we ignore the degeneracy in the parameters in Equation \ref{eq:edacr}, as given by  

\begin{equation}\label{eq:edatxi}
	n(\xi)=K \sqrt{\mathbb{C}} {\xi^{\kappa-1}} \rm{exp}\left[-\frac{\psi}{\kappa}\,{\xi}^{\kappa}\right]
\end{equation}
where 
\begin{equation}\label{eq:psi_kappa_ac}
\psi=\eta_R\left(\frac{1}{\mathbb{C}\gamma_R^2}\right)^{\kappa/2}=\frac{\eta_R}{\xi_{R}^{\kappa}}
\end{equation}
and the normalization K is given below 
\begin{equation}
K = Q_0\,t_{acc,R} \,{{\xi_{R}}^{-\kappa}}\rm{exp}\left[\frac{\eta_R}{\kappa}\,\left(\frac{\xi_0}{\xi_R}\right)^{\kappa}\right]
\end{equation}
Like EDD model, the fit parameters of \emph{$synconv \otimes n(\xi)$} model for energy dependent acceleration scenario are norm $\mathbb{N}$, $\kappa$ and $\psi$. Similarly, $\mathbb{N}$ is defined as
\begin{equation}\label{eq:norm_eda}
\mathbb{N}  = \frac{\delta^3(1+z)}{d_L^2} V \,\mathbb{A}\,K\,\sqrt{\mathbb{C}}
\end{equation}
The plot between the difference of the reduced-$\chi^{2}$ values of the log-parabola and the EDA model vs. reduced-$\chi^{2}$ values of the log-parabola model (shown in Fig. \ref{fig:redchi}(c)) suggests that the EDA model fits the X-ray spectrum equally well as the log-parabola model. Thus confirming the power of the model to reproduce the curvature in the spectrum.  
The Spearman's correlation results between the EDA model fit parameters are presented in the last rows of Table \ref{tab:R}, and the correlation plots are shown in Fig. \ref{fig:eda}. The correlations of flux, $F_{0.3-10 keV}$ with $\kappa$ and $\psi$  are nearly similar to the results obtained in the case of energy dependent $t_{esc}$ model. In this case, the $r_{s}$ ($P_{s}$) are obtained as $0.36\pm0.02$ ( $1.81\times{10^{-24}}$ ) for $\kappa$ vs. $F_{0.3-10 keV}$, and $-0.77\pm0.01$ ( $1.17\times{10^{-181}}$ ) for $\psi$ vs $F_{0.3-10 keV}$ respectively. Nevertheless, $\psi$ and $\kappa$ showed a weak anticorrelation with $r_{s}$ ($P_{s}$) as $-0.41\pm0.01$ ( $2.11\times{10^{-37}}$ ). 
Now, the relation $\psi=\eta_R {\xi_{R}}^{-\kappa}$ implies that $log_{10}\psi$ should be linearly and inversely proportional to $\kappa$, which is consistent with the correlation obtained between $\psi$ and $\kappa$. Fig. \ref{fig:eda}(a) shows $log_{10}\psi$ versus $\kappa$ plot, fitted with a straight line $log_{10}\psi$ = $-0.36\kappa$+0.48. The result implies that $\eta_{R}$ ${\sim}$ 3.02 and ${\xi}_{R}$ ${\sim}$ 2.29 keV. 
The variation of the normalization with $\kappa$ can be represented with the relation given by Equation \ref{eq:norm_eda}, and can be written as 
\begin{equation}\label{eq:log_norm_kappa_eda}
\log \mathbb{N}= \frac{\eta_R}{\kappa}\,A^{\kappa}-{\kappa}\,{log\,{\xi_{R}}}+B
\end{equation}
The log\,$\mathbb{N}$ vs. $\kappa$ plot fitted with the above relation (see Fig. \ref{fig:eda}(b)) resulting the parameter values as A = 0.15, B = 4.10, and ${\eta_R}$ = 3.02, which provides the value of $\gamma_{0}$ $\sim$ 0.15 $\gamma_{R}$. Unlike the EDD model, the photon energy corresponding to $\gamma_{R}$ is ${\xi_{R}}^{2}$ $\sim$ 5.24 keV which is within the energy range used for the spectral study. Furthermore, $\gamma_{0}$ is not notably smaller than $\gamma_{R}$.

\subsection{Comparison between the results presented here and in Hota et al. 2021:}
We compare the correlation study results performed with the long-term observations which include both the flaring and quiescent states, with the ones obtained with short-term flare in \cite{2021MNRAS.tmp.2632H}. The comparison for different models is described below: 
\begin{enumerate}
\item Log-parabola model:
In both the studies, a strong anticorrelation is obtained between $\alpha$ and flux. This shows that the hardening when brightening trend is independent of the state of source. Both the studies showed that $\alpha$ and $\beta$ are inversely correlated and the correlation becomes weaker in the long-term observation involving flaring and quiescent states. Additionally, in case of short-term flare, the value of $\alpha$ decreases slowly with normalization till $\mathbb{N}$ $\sim$ 2 and then decreases more rapidly. While in the present study, $\alpha$ decreases slowly till $\mathbb{N}$ $\sim$ 20, and then it remains constant with $\mathbb{N}$. On the other hand, a positive correlation is observed between $\beta$ and flux in short-term flare, while we found no correlation between $\beta$ and flux in the present study. The critical analysis of the differences and similarities of the results between the short-term and long-term studies are given for the more general physical models, discussed in the subsequent sections. \\

%$\beta$ vs. normalization - in Hota et al. 2019, there is a slow increase of $\beta$ followed by a rapid increase. In the present study, a mild positive correlation implies that there is a slow increase in  $\beta$ with respect to N. 
\item Power-law particle distribution with maximum electron energy model:
The correlation between p and flux followed a similar trend as observed in the correlation of log-parabola
parameters $\alpha$ and flux. Thus further confirming the harder when brighter feature is independent of the flux state  of the source. In \cite{2021MNRAS.tmp.2632H},  p showed little variation with $\mathbb{N}$, till $\mathbb{N}$ $\sim$ 2 and rapidly decreases for higher values, while in the present work parameters p and $\mathbb{N}$ vary rapidly till $\mathbb{N}$ $\sim$ 20, then p remained almost constant for higher values of $\mathbb{N}$. Like p and $\mathbb{N}$, similar behavior is noticed in $\xi_{max}$ vs. $\mathbb{N}$ plot. A moderate anticorrelation is obtained between $\xi_{max}$ and flux in the short-term flare, while, in the present study no correlation is obtained between these quantities. Moreover, p vs. $\xi_{max}$  in a short-term flare is tightly correlated. On the other hand, a weak positive correlation is observed between $\xi_{max}$ and p in the present study. However, we note that the observed correlations are not consistent with the assumptions of the model.
\\
\item EDD model:
In both studies, a strong anticorrelation is observed between $\psi$ and flux. Also,  a strong correlation is obtained between $\kappa$ and flux in the short-term flare, while a low positive correlation is observed in the present study. Therefore, we observe approximately similar correlation results for both the short-term as well as long-term studies. In \cite{2021MNRAS.tmp.2632H}, the variation of $\mathbb{N}$ with $\kappa$ provides an estimate of ${\gamma}_0$ as $\sim$ 0.26 ${\gamma}_R$, while in the present study the value of ${\gamma}_0$ is obtained as $\sim$ 0.045 ${\gamma}_R$. Moreover, in the short-term flare, the photon energy is estimated as ${{\psi}_R}^{2}$ $\sim$ 5.75 keV. The obtained photon energy is within the energy range used, 0.3-10 keV considered for the spectral fitting. While, in the present study, ${{\psi}_R}^{2}$ $\sim$ 10.96 keV is slightly higher than the energy range.  
\\
\item EDA model:
Similar to the case of EDD model, a strong anticorrelation is observed between $\psi$ and flux in both the studies. Also a strong correlation is obtained between $\kappa$ and flux in short-term flare, while a low positive correlation is observed between $\kappa$ and flux in the present study.
In \cite{2021MNRAS.tmp.2632H}, the value  of ${\gamma}_0$ is obtained as $\approx$ 0.19 ${\gamma}_R$, while in the present study ${\gamma}_0$ $\approx$ 0.15 ${\gamma}_R$. In both works, the EDA model provides the estimation of observed photon energy well within the energy range 0.3--10 keV, \cite{2021MNRAS.tmp.2632H}, provides an estimate of photon energy  ${{\psi}_R}^{2}$ $\sim$ 2.89 keV, and in the present study, ${{\psi}_R}^{2}$ $\sim$ 5.24 keV. 
\end{enumerate}

\section{Summary and discussions}\label{sect:disc}
\begin{figure*}

\includegraphics[scale=0.5, angle=-90]{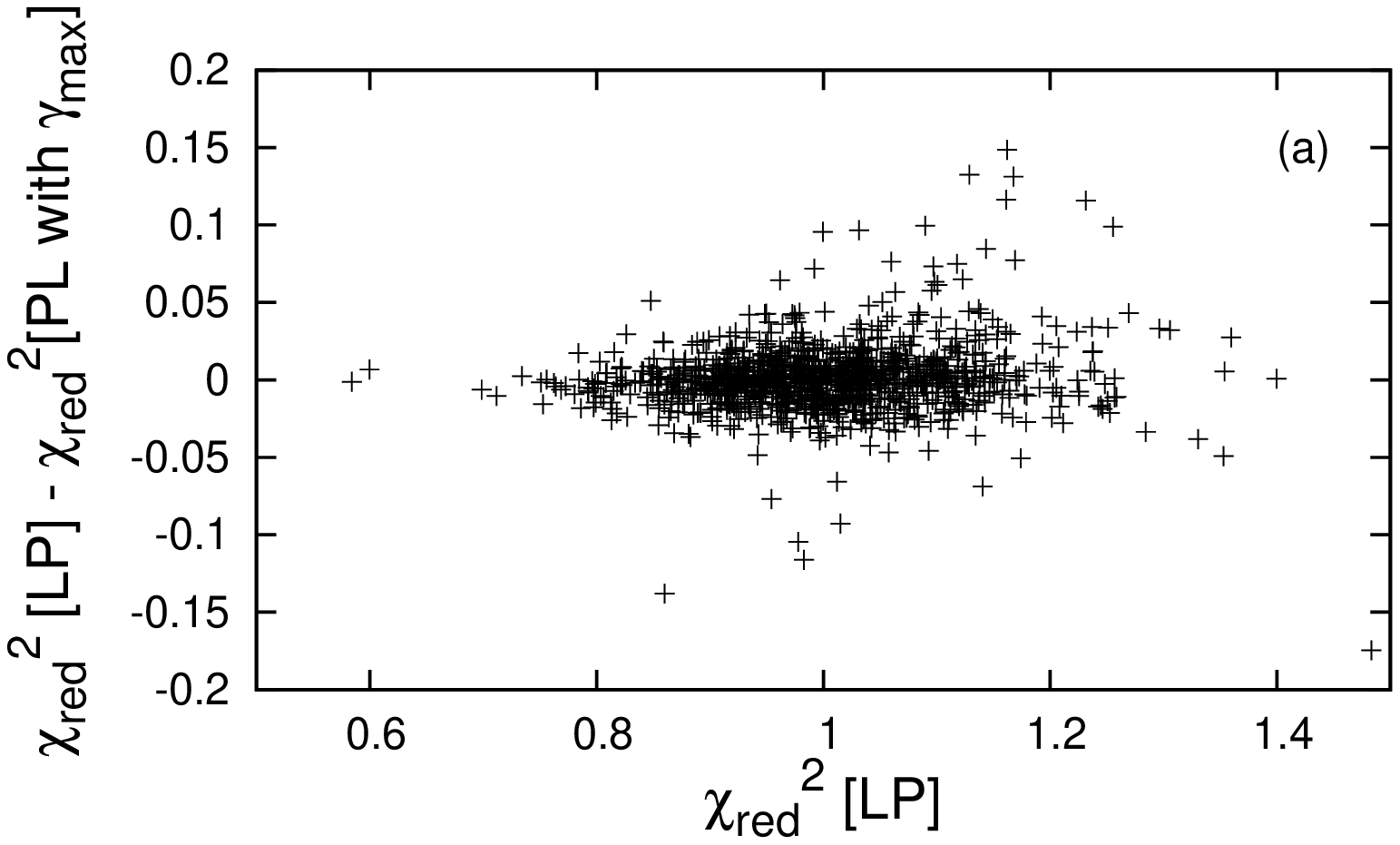}
\includegraphics[scale=0.5, angle=-90]{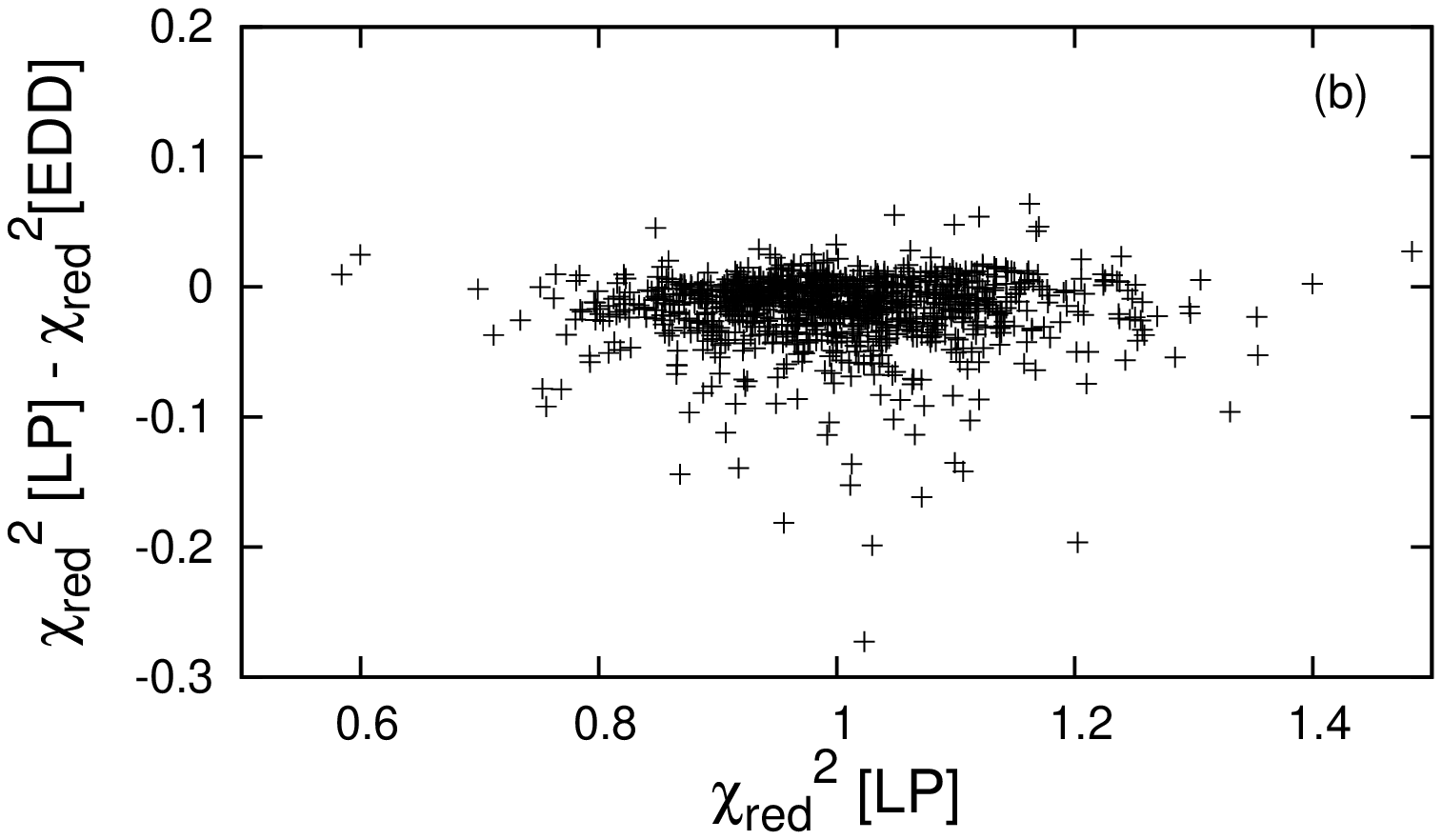}\\
\includegraphics[scale=0.5, angle=-90]{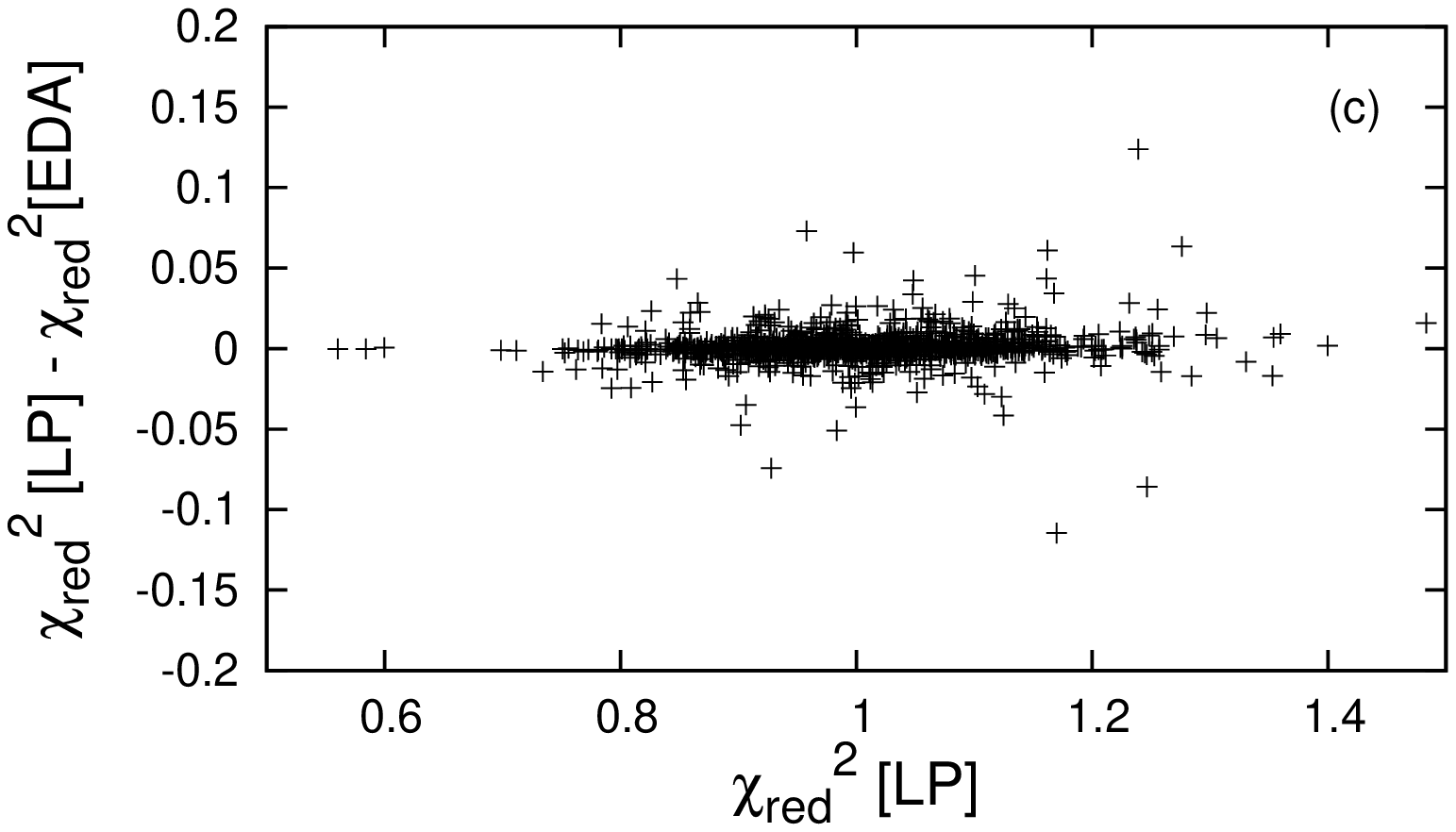}\\
\caption{Panels (a-c): plots for the difference between the reduced-$\chi^{2}$ values of the log-parabola (LP) model to the power-law with $\xi_{max}$, energy-dependent $t_{esc}$ and the energy-dependent $t_{acc}$ models, vs. the LP model.} 
\label{fig:redchi}

\end{figure*}

The X-ray spectrum of Mkn\,421 exhibits a mild curvature during various flux states, and the log-parabola model gives a well fit statistics to such spectrum \citep{2004A&A...413..489M, 2009A&A...504..821P, 10.1093/mnras/stw095, Gaur_2017, 2018ApJ...859...49P}. In this work; in addition to the log-parabola model, we use the physically motivated models which involve acceleration of particles near the shock front and subsequent emission through the synchrotron mechanism. We show that, the convolution of the single-particle synchrotron emissivity with the particle density acquired from physically motivated models can also result curvature in the spectrum. In this regard, each spectrum of Swift-XRT (0.3-10 KeV) observations from April 2005 to April 2020, for which spectral counts > 3000 were fitted with the \emph{SYNCONV} \emph{($synconv \otimes n(\xi)$)} model with $n(\xi)$ represented by log-parabola model, power-law with maximum electron energy, energy-dependent electron diffusion (EDD), and energy-dependent acceleration (EDA) models, respectively. Subsequently, we compared the best fit reduced chi-square (${\chi_{red}}^2$) values of the models to the log-parabola one, as shown in Fig. \ref{fig:redchi}, and found that in addition to the log-parabola model, these models are equally good to fit the spectrum. However, we show that the correlation studies between the best fit model parameters provide important information about the consistency/propriety of the model in representing the observed X-ray spectrum. 

Log-parabola model being a simplified version of more general physical model, we
%In the case of log-parabola model, the relation of $\alpha$ and $\beta$ in terms of the physical parameters being unclear makes it difficult to obtain a physical picture of the emission from the correlation study. Therefore, it becomes important to 
consider the physical models capable of reproducing the curvature.
In the case of PL with maximum electron energy model, we noted a weak positive correlation between the index (p) of the electron distribution and the maximum  Lorentz factor ($\xi_{max}$) of the electron. This result contradicts with the model prediction where the variation in the observed parameters depend on the variation of the acceleration time-scale, which in turn suggests an anticorrelation between p and $\xi_{max}$. Such a positive correlation can occur if it is postulated that the acceleration time-scale  varies inversely with the magnetic field.

We showed that the energy dependent electron diffusion (EDD) model can also reproduce the spectral curvature. While carrying out the correlation study between the model parameters, we found that $\mathbb{N}$ and $\psi$ are strongly anticorrelated with $\kappa$, which is consistent with the model prediction. The appropriateness in the observed correlation with the expected one let us estimate the typical photon energy arising from the electron of energy $\gamma_{R}$, as ${\xi_{R}}^{2} \approx$ 10.96 keV and the injection energy of the electron into the acceleration region, as $\gamma_{0} \approx $ 0.045 $\gamma_{R}$. Similar to the EDD model, EDA model with energy dependent acceleration rate also reproduces the observed X-ray spectrum well and predicts the observed correlation results. In the EDA model, the estimated value of ${\xi_{R}}^{2}$ is $\approx$ 5.24 keV, and $\gamma_{0}$ $\approx$ 0.15 $\gamma_{R}$. Therefore, both the EDD and EDA models provide insight into underlying physical mechanism responsible for X-ray emission. In the earlier work, \citealp{2021MNRAS.tmp.2632H} had shown similar correlation results between the spectral parameters where they investigated the capability of these models to fit the X-ray observations from a single short-term flare. 

 If we consider the case that the energy dependence of the escape or the acceleration time-scale is linked to the gyration radius of the electron, the $\gamma_{R}$ can possibly be the energy where the gyration radius will be equal to the size of the system. In such a case, beyond the energy $\gamma_{R}$, time-scale will become energy independent. Therefore, in accordance with the model compatibility, the value of the characteristic synchrotron photon energy corresponds to the electron energy $\gamma_{R}$ should be larger than the energy range of the observations. The value of ${\xi_{R}}^{2}$ estimated in the EDD model is slightly beyond the observed energy, while for the EDA model it is within the  energy range. Alongside, the injection energy $\gamma_{0}$ of the electron comes out to be 0.045 times $\gamma_{R}$ for the EDD model, which is possibly suitable in the physical framework, compared to the case in the EDA model where $\gamma_{0}$ is 0.15 times $\gamma_{R}$.

We carried out the derivations for the particle distributions by assuming that the radiative cooling is not significant. This is a possible scenario, as it successfully reproduces the observed correlation. If we take the radiative cooling into account, the inferred parameters may be more physical. However in such cases, the form of the particle energy distributions will not be  analytical  and hence one has to undertake  numerical techniques to compute and fit the data. Therefore the models we have used here can be further modified to represent more realistically by considering the following scenario/assumptions. As already pointed out in \cite{2021MNRAS.tmp.2632H} that the energy dependence of the diffusion and the acceleration time-scales from power-law into a complex form could modify the electron distribution and hence the correlation results and the values of the physical parameters. Further, a more physical scenario would be to consider both the escape and acceleration time-scales to be energy dependent. A connection between the energy-dependent acceleration and /or escape time-scale with the momentum diffusion coefficient has been investigated by various authors \citep{Stawarz_2008, 2011ApJ...739...66T, 10.1093/mnras/stu1060}. Moreover, the inclusion of stochastic acceleration ( or the second-order Fermi acceleration) and the presence of momentum-diffusion term may refine the particle distribution profile \citep{1962SvA.....6..317K, 2007A&A...467..501T, 2009A&A...501..879T, 2011ApJ...739...66T}. Another important point can be, while the observed correlation between the particle index and the maximum Lorentz factor is explained when the magnetic field variation is associated with the time-scale, it can be investigated whether magnetic field controls the acceleration process.  

In our work, we have probed the spectral feature of Mkn\,421 in the energy range 0.3--10 keV using Swift-XRT observations, however, an extended study to hard X-ray is required to understand the role of the considered models better. 

%######################

%%%%% USING MONTE CARLO ANALYSIS
\begin{table*}
\centering
\caption{Spearman Correlation results of the observations with spectral counts > 3000}
\begin{tabular}{lccc}
\hline
\hline
Model   &  \multicolumn{1}{c}{Correlation between }    &  \multicolumn{1}{c}{${r_{s}}$} &   \multicolumn{1}{c}{${P_{s}}$}\\ 

 \hline

Log-parabola & $\alpha$ \&  Flux & $-0.81 \pm 0.005$ & $4.67\times10^{-229}$ \\
& $\beta$ \& Flux & $0.04\pm 0.02$ &  0.27 \\
& $\alpha$ \&  $\mathbb{N}$ & $-0.41 \pm 0.008$ & $1.78\times10^{-39}$ \\
& $\beta$ \&  $\mathbb{N}$ & $0.15 \pm 0.02$ & $2.37\times10^{-05}$ \\
& $\alpha$ \& $\beta$ & $-0.21 \pm 0.02$ & $7.68\times10^{-09}$\\

\hline
PL with $\xi_{max}$ & p \& Flux & $-0.80 \pm 0.007$ & $1.92\times10^{-209}$ \\
& ${\xi_{max}}$ \& Flux & $-0.03 \pm 0.02$ &  $0.44$ \\
& p \& $\mathbb{N}$ & $-0.28 \pm 0.009$ & $6.31\times10^{-19}$ \\
& ${\xi_{max}}$ \& $\mathbb{N}$ & $-0.15 \pm 0.02$ &  $2.28\times10^{-4}$ \\
& ${\xi_{max}}$ \& p & $0.20 \pm 0.02$ &  $4.94\times10^{-7}$\\

\hline

Energy-dependent $t_{esc}$ & ${\kappa}$ \&  Flux & $0.32 \pm 0.02$ & $3.30\times10^{-19}$ \\
& ${\psi}$ \& Flux & $-0.76 \pm 0.008$ & $1.01\times10^{-180}$ \\
& ${\kappa}$ \&  $\mathbb{N}$ & $-0.68 \pm 0.02$ & $7.95\times10^{-114}$ \\
& ${\psi}$ \& $\mathbb{N}$ & $0.70 \pm 0.01$ & $2.05\times10^{-121}$ \\
& ${\psi}$ \& ${\kappa}$ & $-0.51 \pm 0.01$ & $3.65\times10^{-51}$ \\

\hline

Energy-dependent $t_{acc}$ & ${\kappa}$ \&  Flux & $0.36 \pm 0.02$ & $1.81\times10^{-24}$ \\
& ${\psi}$ \& Flux & $-0.77 \pm 0.01$ & $1.17\times10^{-181}$ \\
& ${\kappa}$ \&  $\mathbb{N}$ & $-0.78 \pm 0.009$ & $9.17\times10^{-194}$ \\
& ${\psi}$ \& $\mathbb{N}$ & $0.73 \pm 0.004$ & $4.95\times10^{-170}$ \\
& ${\psi}$ \& ${\kappa}$ & $-0.41 \pm 0.01$ & $2.11\times10^{-37}$ \\				   
	 \hline    
    
        \end{tabular}
       \label{tab:R}
        						   
\end{table*}

%%%%%%%%5 figures

%%%%%%%%%%%%%%

%########################## tab1

%######################################### 
\section*{Acknowledgements}
We thank the anonymous reviewer for valuable comments and suggestions. We acknowledge the use of public data from the Swift data archive. This work made use of data supplied by the UK Swift Science Data Centre at the University of Leicester. R. Khatoon and R. Gogoi would like to thank CSIR, New Delhi (03(1412)/17/EMR-II) for financial support. R. Gogoi would like to thank IUCAA, Pune for associateship. Z. Shah is supported by the Department of Science and Technology, Govt. of India, under the INSPIRE Faculty grant
(DST/INSPIRE/04/2020/002319).
%###########################
\section*{Data Availability}

The data and softwares used in this research are available at NASA's HEASARC webpages with the links given in the manuscript. The table with best-fitting parameters of spectral analysis fitted with the synchrotron convolved Log-parabola, Power-law particle distribution with maximum electron energy model, EDD model and EDA model is made available online.

\bibliographystyle{mnras}
\bibliography{references} % if your bibtex file is called example.bib

%%%%%%%%%%%%%%%%% APPENDICES %%%%%%%%%%%%%%%%%%%%%

% Don't change these lines
\bsp	% typesetting comment
\label{lastpage}
\end{document}